\documentclass[11pt]{article}

\usepackage{stolenstyle}

\usepackage{amsmath,epsf,amssymb,latexsym,amsthm,setspace,bbm,array,pifont,multirow}

\usepackage[matrix,arrow,frame,import,curve,color]{xy}

\DeclareFontFamily{U}{rsf}{}
\DeclareFontShape{U}{rsf}{m}{n}{
  <5> <6> rsfs5 <7> <8> <9> rsfs7 <10-> rsfs10}{}
\DeclareMathAlphabet\Scr{U}{rsf}{m}{n}

\def\rep#1{{{\boldsymbol{#1}}}}

\def\cDb{{\overline{\cD}}}

\def\P{{\mathbb P}}
\def\R{{\mathbb R}}
\def\Z{{\mathbb Z}}

\def\rank{\operatorname{rank}}

\def\SO{\operatorname{SO}}

\def\SU{\operatorname{SU}}
\def\GU{\operatorname{U{}}}

\def\GE{\operatorname{E}}

\def\so{\operatorname{\mathfrak{so}}}

\def\su{\operatorname{\mathfrak{su}}}

\def\ff#1#2{{\textstyle\frac{#1}{#2}}}
\def\half{\frac{1}{2}}

\def\cD{{\cal D}}
\def\cE{{\cal E}}

\def\cL{{\cal L}}

\def\cO{{\cal O}}

\def\cV{{\cal V}}
\def\cW{{\cal W}}

\newcommand\alphab{\overline{\alpha}}

\newcommand\mub{\overline{\mu}}

\newcommand\chib{\overline{\chi}}

\newcommand\Gammah{\widehat{\Gamma}}

\newcommand\Gammat{\widetilde{\Gamma}}

\newcommand\bb{\overline{b}}
\newcommand\cb{\overline{c}}

\newcommand\qb{\overline{q}}
\newcommand\rb{\overline{r}}

\newcommand\Jh{\widehat{J}}

\newcommand\Gb{\overline{G}}

\newcommand\Jb{\overline{J}}

\newcommand\Tb{\overline{T}}

\newcommand\Bt{\widetilde{B}}

\newcommand\Jt{\widetilde{J}}

\newcommand\St{\widetilde{S}}
\newcommand\Tt{\widetilde{T}}

\makeatletter
\newsavebox{\@brx}
\newcommand{\llangle}[1][]{\savebox{\@brx}{\(\m@th{#1\langle}\)}%
  \mathopen{\copy\@brx\kern-0.5\wd\@brx\usebox{\@brx}}}
\newcommand{\rrangle}[1][]{\savebox{\@brx}{\(\m@th{#1\rangle}\)}%
  \mathclose{\copy\@brx\kern-0.5\wd\@brx\usebox{\@brx}}}
\makeatother

\def\bJ{{{\boldsymbol{J}}}}

\def\bY{{{\boldsymbol{Y}}}}

\def\bdelta{{{\boldsymbol{\delta}}}}
\def\bgamma{{{\boldsymbol{\gamma}}}}
\def\blambda{{{\boldsymbol{\lambda}}}}
\def\balpha{{{\boldsymbol{\alpha}}}}
\def\balphab{{{\boldsymbol{\alphab}}}}

\def\bxi{{{\boldsymbol{\xi}}}}
\def\bmu{{{\boldsymbol{\mu}}}}
\def\bmub{{\boldsymbol{\mub}}}
\def\bnu{{{\boldsymbol{\nu}}}}
\def\br{{{\boldsymbol{r}}}}

\def\bq{{{{\boldsymbol{q}}}}}

\def\by{{{{\boldsymbol{y}}}}}
\def\bbb{{{{\boldsymbol{b}}}}}
\def\bB{{{{\boldsymbol{B}}}}}
\def\bBt{{{{\boldsymbol{\Bt}}}}}
\def\bqb{{{\overline{\bq}}}}
\def\sb{{{\overline{s}}}}

\def\Gammah{{{\widehat{\Gamma}}}}
\def\Gammat{{{\widetilde{\Gamma}}}}

\def\tot{{{\text{tot}}}}

\def\PF{{{\text{PF}}}}
\def\GP{{{\text{GP}}}}

\def\mod{{{\text{ mod }}}}

\def\preprint{IPMU18-0189}

\title{A (0,2) mirror duality}
\author[a] {Marco Bertolini}
\author[b] {and M.~Ronen Plesser}
\affiliation[a]{Kavli Institute for the Physics and Mathematics of the Universe (WPI),\\
The University of Tokyo, Kashiwa, Chiba 277-8583, Japan}
\affiliation[b]{Center for Geometry and Theoretical Physics, Box 90318 \\
Duke University, Durham, NC 27708-0318, USA}
\emailAdd{marco.bertolini@ipmu.jp}
\emailAdd{plesser@cgtp.duke.edu}

\abstract{We construct a class of exactly solved (0,2) heterotic
compactifications, similar to the (2,2) models constructed by
Gepner.  We identify these as special points in moduli spaces
containing geometric limits described by non-linear sigma models on
complete intersection Calabi--Yau spaces in toric varieties,
equipped with a bundle whose rank is strictly greater than that of
the tangent bundle.  These moduli spaces do not in general contain a
locus exhibiting (2,2) supersymmetry.  A quotient procedure at the
exactly solved point realizes the mirror isomorphism, as was the
case for Gepner models.  We find a geometric interpretation of the
mirror duality in the context of hybrid models.}

\begin{document}

\maketitle

\section{Introduction}
\label{s:intro}
 
Exactly solvable conformal field theories (CFTs) \cite{Gepner:1987vz} have
played a crucial role in unraveling the structure of the corresponding moduli spaces.
Perhaps the most striking
example, and surely the most relevant to the present discussion, is
mirror symmetry, which was first discovered as a duality of exactly
solvable (2,2) superconformal field theories (SCFTs) describing a
(2,2) non-linear sigma model (NLSM) on a Calabi-Yau (CY) target space
\cite{Greene:1990ud}.  The dual theory, obtained by a quotient
procedure leading to an isomorphic SCFT, corresponds to a point in the
moduli space of a topologically distinct CY space. Using
superconformal perturbation theory the two moduli spaces are thus
identified through the mirror map, and the isomorphism relates
computations in one theory to computations in the other.
Subsequently, these were found to be special cases of a larger class
of models described by Abelian gauged linear sigma models (GLSMs)
\cite{Witten:1993yc}, for which the duality and the mirror map
\cite{Aspinwall:1993rj} have a natural combinatorial description
\cite{Morrison:1995yh}.  When these admit a geometric description as
an NLSM the combinatorial duality relates pairs of complete
intersection CY (CICYs) spaces in toric varieties \cite{Batyrev:1994pg}.
Used as the internal space for a
compactification of the heterotic string, mirror theories lead to
theories related by charge conjugation in four dimensions.

SCFTs with (2,2) supersymmetry have deformations preserving a (0,2)
subalgebra, extending the space of heterotic compactifications.  The
mirror isomorphism extends to these, raising the question: to what
extent does mirror symmetry extend to more general (0,2) SCFTs?
Despite some early but sporadic evidence \cite{Blumenhagen:1996vu},
almost the entirety of subsequent studies has focused on theories
which are obtained by deformations of (2,2) models
\cite{Melnikov:2012hk}. Even in this very restricted set of theories
novel challenges arise. 
For instance, (0,2) deformations of 
(2,2) GLSMs are not in general mirror symmetric \cite{Kreuzer:2010ph}, and the associated B/2-twisted theory 
is not protected in principle by worldsheet instanton corrections \cite{McOrist:2008ji}. However, these issues have been at least partially addressed:
by restricting to a subset of such deformations, mirror symmetry is restored \cite{Melnikov:2010sa,Bertolini:2018qlc},
and for a large class of theories, the B/2 model appers to be insensitive to quantum corrections \cite{Bertolini:2018qlc}.
Hence, from both a conceptual and a computational point of view mirror
symmetry extends reasonably well outside the (2,2) locus, and one
might be prompted to explore whether this duality extends to more
general (0,2) theories.  

In this work, we begin an exploration of the general question raised
above.  In particular, we describe a class of (0,2) SCFTs not related
to deformations of (2,2) models and show that this class exhibits
mirror symmetry.  Employing the recent results of
\cite{Gadde:2016khg}, we construct exactly solvable (0,2) SCFTs
describing a point in the moduli space of $N=1$ heterotic string vacua
adapting the ideas of \cite{Gepner:1987vz}.  We then show that an
orbifold procedure, generalizing the construction of
\cite{Greene:1990ud} for Gepner models, yields an isomorphic conformal
field theory which differs from the original theory in the sign
of the right-moving $\GU(1)$
R-charge.\footnote{Numerological evidence that an orbifold procedure
  might extend mirror symmetry to (0,2) models dates back to
  \cite{Blumenhagen:1996tv}.}

Exactly as for (2,2) mirror symmetry, the duality at this point extends 
via superconformal perturbation theory
to the whole moduli space.
It is therefore natural to ask whether there is a geometric
counterpart to our mirror procedure.  The answer turns out to be
affirmative if we are willing to generalize our notion of geometry. 
A key role in utilizing the
symmetry in the (2,2) case was the realization \cite{Candelas:1990qd}
that there is a subspace of the moduli space, extending from the
Gepner point all the way to the large-radius limit in which the NLSM
is a good description, preserving the discrete symmetry used to
construct the dual.  Thus the mirror CY could be described as a
quotient of the original target space. The GLSM provided a framework
for understanding this deformation and a more general setting for the
duality.  Underlying this is the factorization of the moduli space of
(2,2) SCFTs as a local product, which is absent in (0,2) moduli
spaces.  A novel issue that arises in the (0,2) setting is that for the class of
models under study in this work, although we have a geometric
interpretation in terms of a NLSM on
$\cV\rightarrow M$, where $M$ is a CICY and $\cV$ is a stable
holomorphic vector bundle, it turns out that there is no smooth
geometry preserving the discrete symmetry.
In fact, from the point of view of the geometry $\cV\rightarrow M$,
the enhanced symmetry locus  
obtained by realizing the model in a (0,2) GLSM and tuning the FI
term to obtain a large radius description 
while keeping all the other moduli fixed leads to a singular complex
structure for $M$.  We will show that not all is lost however. 
This locus of singular complex structures admits an alternative description in terms of a (0,2) hybrid model \cite{Bertolini:2017lcz}. 
Briefly, such models are defined as NLSMs on the geometry $\cE\rightarrow \bY$,
where the target space $\bY$ is generically non-compact and not
necessarily Calabi-Yau, equipped with a (0,2) superpotential
interaction $\bJ\in\Gamma(\cE^\ast)$ such that the space of classical vacua is a compact
K\"ahler subspace $\bJ^{-1}(0)=B\subset \bY$.
Thus, for the class of models we consider, mirror symmetry has a natural interpretation
in terms of hybrid models and quotients thereof.

In a smooth geometric limit, we can label deformations, to some
approximation, as associated to the complex structure deformations of
$M$, deformations of the complexified K\"ahler class, and deformations
of the holomorphic bundle $\cV$.  In a hybrid limit, we have a similar
description in terms of the complexified K\"ahler class of $B$, complex structure
deformations of $\cE\rightarrow\bY$, and deformations of the superpotential
interaction.  The quotient construction provides some hints as to how
the duality acts on these.  In some cases, as we shall see, the quotient
acts on $B$ only.  This imposes invariance constraints on the
deformations of complex structure of $\cE\rightarrow\bY$ and of the superpotential, decreasing the
dimension of these spaces.  Resolving the quotient singularities of
$B$ will introduce new K\"ahler moduli for the base, as well as
possibly new deformations associated to the extension of the bundle
and of the superpotential interaction over the exceptional divisor.  
This heuristically echoes the way a quotient in the (2,2) case removes
some complex structure deformations by imposing an invariance
constraint and introduces new K\"ahler deformations when quotient
singularities are resolved.

The rest of this paper is organized as follows. In section \ref{s:GPmodels} we review, following \cite{Gadde:2016khg},
the exact solution of a certain class of (0,2) Landau-Ginzburg theories, and we further study some of its properties. 
In section \ref{s:exactstrings} we derive, using such models as part of the construction,
a modular invariant partition function describing a consistent
heterotic string theory vacuum following \cite{Gepner:1987vz}.
We show that an orbifold procedure, generalizing the Greene-Plesser construction, yields an isomorphic conformal field theory
where all the states have reversed right-moving R-charge. 
We also give a geometric interpretation of such a mirror quotient, and show that it is more naturally interpreted in terms of quotients of hybrid models rather 
than compact Calabi-Yau manifolds with holomorphic bundles.  
In section \ref{s:examples} we present some explicit examples elucidating our construction, and we conclude in section
\ref{s:outlook} with some prospectives that we believe are naturally opened by the existence of such a mirror duality.

\acknowledgments

It is a pleasure to thank M.~Romo for important contributions at the
inception and early stages of this project, and P.S.~Aspinwall, I.V.~Melnikov and
D.R.~Morrison for discussions.  
MRP thanks the CERN theory group for their kind hospitality during the final
stages of this work.
This work was supported by World Premier International Research Center
Initiative (WPI Initiative), MEXT, Japan (MB), and by NSF Grant
PHY-1521053 (MRP).
Any opinions, findings, and conclusions or
recommendations expressed in this material are those of the authors
and do not necessarily reflect the views of the National Science
Foundation.

\section{GP${}_{m,n}$ models}
\label{s:GPmodels}
 
In this section we review the proposal of \cite{Gadde:2016khg} 
concerning the exactly solvable structure of conformal field theories 
which admit a UV free realization in terms of 
a certain class of (0,2) Landau-Ginzburg models \cite{Distler:1993mk,Kawai:1994np,Melnikov:2009nh}.
Let us consider the class of LG theories described by the (0,2) superpotential
\begin{align}
\label{eq:02LGsupnm}
\cW= \Gamma^1 \Phi_1^n + \Gamma^2\Phi_2^m + \Gamma^3 \Phi_1\Phi_2~.
\end{align}
This theory exhibits, in addition to the naive right-moving R-symmetry $\GU(1)_R^0$, a global $\GU(1)^2$ symmetry, given by
\begin{align}
\xymatrix@R=0mm@C=5mm{
\text{fields}	&\Phi_1		&\Phi_2		&\Gamma^1		&\Gamma^2		&\Gamma^3			&\theta\\
\GU(1)_R^0	&0			&0			&1				&1				&1					&1\\
\GU(1)_1		&1			&0			&-n				&0				&-1					&0\\
\GU(1)_2		&0			&1			&0				&-m				&-1					&0
}
\end{align}
A simple application of $c$-extremization \cite{Benini:2012cz,Benini:2013cda} yields the following values for the charges of the fields under the 
right-moving R-symmetry of the IR CFT
\begin{align}
\label{eq:RsymGPmn}
\xymatrix@R=0mm@C=5mm{
\text{fields}	&\Phi_1		&\Phi_2		&\Gamma^1		&\Gamma^2		&\Gamma^3\\
\GU(1)_R		&\ff{m}{mn+1}	&\ff{n}{mn+1}	&\ff{1}{mn+1}		&\ff{1}{mn+1}		&-\frac{m+n}{mn+1}+1
}
\end{align}
These, in particular, determine the data
\begin{align}
\label{eq:cftdata}
c&=3\frac{mn-1}{mn+1}+1~,		&\cb&=3\frac{mn-1}{mn+1}~,		&r&=\frac{2mn}{mn+1}~.
\end{align}
The $\mathfrak{u}(1)_L$ level-$r$ Kac-Moody (KM) algebra defined by the linear combination
\begin{align}
\label{eq:U1LGPmn}
\GU(1)_L\equiv \GU(1)_R-\GU(1)_R^0 = \frac1{mn+1}\left( m\GU(1)_1+n\GU(1)_2 \right)
\end{align}
plays a fundamental role in our applications. From the LG perspective, 
we consider 
\eqref{eq:02LGsupnm} 
as a specific form of a more general superpotential preserving 
$\GU(1)_L\times\GU(1)_R$. 
We note that \eqref{eq:U1LGPmn} is the only linear combination of the global symmetries $\GU(1)_{1,2}$ which commutes with $\cDb$-cohomology, where
$\cDb$ is the anti-chiral superderivative. This property allows us to define the notion of finite dimensional topological subrings of 
the infinite dimensional chiral ring of the (0,2) CFT
\cite{Adams:2005tc}, generalizing the (c,c) and (a,c) rings of (2,2)
theories \cite{Lerche:1989uy}.

If $m$ or $n$ is 1 the model is somewhat trivial.
In fact, 
suppose without loss of generality that $n=1$.
Then $\Gamma^3$ has the charge of a free field, and we can perform the field redefinition
\begin{align}
\Gamma'^1&=\Gamma^1+\Phi_2\Gamma^3~,
\end{align}
which leads to the superpotential
\begin{align}
\cW'= \Gamma'^1 \Phi_1 + \Gamma^2\Phi_2^m + \Gamma^3 \times 0~.
\end{align}
The resulting theory takes the form of a product of a free left-moving Fermi field, a massive pair $(\Phi_1,\Gamma'^1)$ 
and a (2,2) minimal model $(\Phi_2,\Gamma^2)$. 
Hence, in the following we will restrict ourselves to the case $n,m\geq2$. 
In this case, it is easy to exclude accidental IR symmetries
\cite{Bertolini:2014ela} not involving one of the Fermi fields
becoming free, and the exact solution suggests free fermion accidents
are absent as well.

In \cite{Gadde:2016khg} it is proposed that the theory defined by \eqref{eq:02LGsupnm} flows in the IR
to the product theory
\begin{align}
\label{eq:IRtheorymn}
\text{GP}_{m,n}=\left(\PF_{mn-1}\times \GU(1)^2_Q\right)\times\overline{C_{mn-1}}~.
\end{align}
That is, the left-moving CFT takes the form of a product of the $\Z_{k}$ parafermion theory $\PF_{k}$ \cite{Fateev:1985mm} and a 
left-moving $\GU(1)^2$ current with anomaly matrix given by
\begin{align}
Q=\begin{pmatrix}
m^2	&1\\
1	&n^2
\end{pmatrix}~.
\end{align}
The right-moving SCFT is an anti-holomorphic level-$k$ $N=2$ minimal
model $\overline{C_{k}}$.\footnote{This is a product in the sense that
states fall into unitary representations of the left-moving algebra
given by the parafermion theory and the Kac-Moody current algebra,
and of the right-moving superconformal algebra with central charge
$\cb$, paired so as to form a closed OPE algebra.}

In the remainder of this section we will 
compute the full partition function of the theory 
following \cite{Gadde:2016khg}, 
and study its modular properties. We note at the outset that the
gravitational anomaly $c-\cb$ precludes modular invariance.  Under
the transformation $T$ the best we can achieve is 
\begin{align}
\label{eq:modcov}
TZ &= e^{-2\pi i\frac{1}{24}}Z~,		&SZ&= Z~,
\end{align}
but it turns out that these conditions are too stringent for our purposes. 
We will in fact construct the full set of pairings of left- and
right-moving characters in this model satisfying \eqref{eq:modcov} up to extra (not necessarily overall) phase factors,
which we will denote ``modular covariant'' partition functions.
We will show that these come in pairs of isomorphic theories, related
by the reversal of the $\GU(1)_R$ charges of all states.

\subsection{The partition functions}

Here we provide the partition function of the theory \eqref{eq:IRtheorymn} in all sectors, defined by the periodicity of the boundary conditions 
for the left- and right-moving fields.
In what follows we will often refer to the R (NS) sector, to be defined by (anti-)periodic boundary conditions for the anti-holomorphic fermions. 

The authors of \cite{Gadde:2016khg} 
compute a partition function for the theory in the NS sector given by
\begin{align}
\label{eq:AApartfunct}
Z_{\text{AA}}=\half \sum_{\alpha=0}^{mn-1} \sum_{\nu\in\Z_{2(mn-1)}}\sum_{t\in\Z_2}\sum_{\sb=0,2}\sum_{a\in \Z_{mn+1}} \chi^{\text{PF}_{mn-1}}_{\alpha,\nu} \chi^{\GU(1)_Q^2}_{(-ma,-n(a+\nu)),t} 
\chib^{\alpha}_{2a+\nu,\sb}~.
\end{align}
where
\begin{align}
\begin{pmatrix}
-ma,-n(a+\nu) 
\end{pmatrix}\in\Z^2_Q\equiv \frac{\Z^2}{Q\Z^2}~.
\end{align} 
We collect 
our conventions for
the various characters appearing here and in what follows
in appendix \ref{app:charconvs}.
The partition function \eqref{eq:AApartfunct} exhibits the following modular transformations 
\begin{align}
T^2 Z_{\text{AA}}&=e^{-2\pi i\frac{2}{24}}Z_{\text{AA}}~,		&S Z_{\text{AA}}&=Z_{\text{AA}}~.
\end{align}

It is not modular covariant since it is not covariant under $T$; this
is expected, as our notation suggests $Z_{\text{AA}}$ imposes
anti-periodic boundary conditions in the time direction as well as in
the space direction.  To obtain a modular covariant combination under
$T$ we can simply construct $\half(Z + TZ)$.  As usual, this
implements a projection onto states invariant under a $\Z_2$ symmetry
which will act as fermion number on right-movers.  We can find this as
follows. 
\eqref{eq:AApartfunct} pairs left/right-moving states with $\GU(1)_L\times\GU(1)_R$ charges $(\bq,\bqb)$ given by 
\begin{align}
\label{eq:NScharges}
\bq&=-\frac{mn(2a+\nu)}{mn+1}+t+2\Z~,		
&\bqb&=\frac{2a+\nu}{mn+1}+\frac{\sb}2+2\Z~.
\end{align}
The charges \eqref{eq:NScharges} satisfy $\bq-\bqb\in\Z$, and denoting $J_0$($\Jb_0$) the current for the $\GU(1)_L$($\GU(1)_R$) symmetry, 
it follows that 
\begin{align}
\label{eq:NSfermnumb}
F_{\text{NS}}=(-1)^{J_0-\Jb_0}=(-1)^{\nu+t-\frac{\sb}2}
\end{align}
is a $\Z_2$ symmetry acting as fermion
number on the right-movers.
Inserting this produces
\begin{align}
\label{eq:APpartfunct}
Z_{\text{PA}}=\half \sum_{\alpha=0}^{mn-1} \sum_{\nu\in\Z_{2(mn-1)}}\sum_{t\in\Z_2}\sum_{\sb=0,2}\sum_{a\in \Z_{mn+1}} (-1)^{\nu+t-\frac{\sb}2} \chi^{\text{PF}_{mn-1}}_{\alpha,\nu} \chi^{\GU(1)_Q^2}_{(-ma,-n(a+\nu)),t} 
\chib^{\alpha}_{2a+\nu,\sb}~.
\end{align}
It is easy to check that this transforms under $T$ as expected, that is, 
\begin{align}
TZ_{\text{AA}}&=e^{-2\pi i\frac{1}{24}}Z_{\text{PA}}~,		&TZ_{\text{PA}}&=e^{-2\pi i\frac{1}{24}}Z_{\text{AA}}~,
\end{align}
verifying that we have correctly identified $(-1)^F$.

As usual, preserving modular invariance will require a twisted sector,
in which the fermions are periodic, namely the R sector. This is
obtained by considering $\sb=\pm1$. 
We begin by defining the AP partition function
\begin{align}
\label{eq:PApartfunct}
Z_{\text{AP}}&=\frac1{2} \sum_{\alpha=0}^{mn-1} \sum_{\nu\in Z_{2(mn-1)}}\sum_{a\in\Z_{mn+1}}\sum_{\sb=\pm1}
\chi^{\text{PF}_{mn-1}}_{\alpha,\nu}\chi^{\GU(1)^2_Q}_{\left(-m(a-\half),-n(a+\nu-\frac{1}2)\right)} \chib_{2a+\nu-1,\sb}^{\alpha}~.
\end{align}
We can check that this is the right object by looking at its modular transformations
\begin{align}
T Z_{\text{AP}}&= e^{-2\pi i\left(\frac1{24}-\frac18\right)}Z_{\text{AP}}~,		&SZ_{\text{PA}}&=Z_{\text{AP}}~,		&SZ_{\text{AP}}&=Z_{\text{PA}}~.
\end{align}
The extra factor in the $T$ transformation of $Z_{\text{AP}}$ is due to the fact that 
lowest-weight left-moving states have weight $r/8\neq c/24$, which gives a contribution 
\begin{align}
\frac{r}8-\frac{\cb}{24}=\frac18\left(\frac{2mn}{mn+1}-\frac{nm-1}{mn+1} \right)=\frac18~.
\end{align}
The $\GU(1)_L$ and $\GU(1)_R$ charges of the states defined in \eqref{eq:PApartfunct} now take the form
\begin{align}
\bq&=-\frac{mn(2a+\nu-1)}{mn+1}+t+2\Z~,	
&\bqb&=\frac{2a+\nu-1}{mn+1}+\frac{\sb}2+2\Z~,
\end{align}
and thus satisfy
\begin{align}
\bq-\bqb&=\nu-1+t-\frac{\sb}2+2\Z~.
\end{align}
Naively, one can try to construct a R sector fermion number analogous to \eqref{eq:NSfermnumb} as
\begin{align}
\label{eq:FFRamond}
F_{\text{R}}=(-1)^{J_0-\Jb_0}=(-1)^{\nu-1+t-\frac{\sb}2}=e^{-\frac{\pi i}2}(-1)^{\nu'+t-\frac{\sb+1}2}~.
\end{align}
In particular, $iF_\text{R}$ does behave as a well-defined fermion
number, the shift again due to the charge of the R ground state, and 
it allows us to determine the last contribution to the partition
function 
\begin{align}
\label{eq:PPpartfunct}
Z_{\text{PP}}&=\frac1{2} \sum_{\alpha=0}^{mn-1} \sum_{\nu\in Z_{2(mn-1)}}\sum_{a\in\Z_{mn+1}}\sum_{\sb=\pm1} (-1)^{\nu'+t-\frac{\sb+1}2}
\chi^{\text{PF}_{mn-1}}_{\alpha,\nu}\chi^{\GU(1)^2_Q}_{\left(-m(a-\half),-n(a+\nu-\frac{1}2)\right)} \chib_{2a+\nu-1,\sb}^{\alpha}~,
\end{align}
which exhibits the following modular transformations
\begin{align}
TZ_{\text{PP}}&=e^{-2\pi i\frac{1}{24}}e^{2\pi i \frac18}Z_{\text{PP}}~,		&SZ_{\text{PP}}&=e^{-2\pi i \frac14}Z_{\text{PP}}~.
\end{align}
A modular covariant partition function of the full theory is obtained by taking the sum over the various sectors of the partition functions constructed above. 
Putting all together we obtain
\begin{align}
Z&=\frac12\left(
Z_{\text{AA}}+Z_{\text{AP}}+Z_{\text{PA}}+Z_{\text{PP}}
\right)\nonumber\\
&=\frac14 \sum_{\alpha=0}^{mn-1} \sum_{\nu\in\Z_{2(mn-1)}}\sum_{t\in\Z_2}\sum_{a\in \Z_{mn+1}}\sum_{b=0}^1 
\chi^{\text{PF}_{mn-1}}_{\alpha,\nu} \chi^{\GU(1)_Q^2}_{(-m(a-\frac{b}2),-n(a-\frac{b}2+\nu)),t} \chib^{\alpha}_{2a-b+\nu,2(\nu+t)-b}~.
\end{align}
A simple change of variables
\begin{align}
q&=2a+\nu-b~,		&s&=2(t+\nu)-b~,
\end{align}
where it is easy to check that $q\in\Z_{2(mn+1)}$ and $s\in\Z_{4}$, leads to the more compact form
\begin{align}
Z&=\frac14 \sum_{\alpha=0}^{mn-1} \sum_{\nu\in\Z_{2(mn-1)}}\sum_{s\in\Z_4}\sum_{q\in \Z_{2(mn+1)}} 
\chi^\alpha_{q,s;\nu} \chib^{\alpha}_{q,s}~,
\end{align}
where we introduced 
\begin{align}
\label{eq:LMcharct}
\chi^\alpha_{q,s;\nu}\equiv\chi^{\text{PF}_{mn-1}}_{\alpha,\nu} \chi^{\GU(1)_Q^{2}}_{(-\frac{m}2(q-\nu),-\frac{n}2(q+\nu)),\frac{s+b}2-\nu}~,
\end{align}
and $b=s\mod2\in\{0,1\}$.
These characters exhibit nice transformation properties. Combining \eqref{eq:PFchtransf} and \eqref{eq:U12chtransf} one can show that
\begin{align}
\label{eq:LMtransfs}
T\chi^\alpha_{q,s;\nu}&=e^{-2\pi i \frac{c_{\text{PF}_{mn+1}}+2}{24}} 
e^{2\pi i \left(\frac{\alpha(\alpha+2)}{4(mn+1)}-\frac{q^2}{4(mn+1)}+\frac{s^2}8+\frac{b^2}8 \right)}\chi^\alpha_{q,s;\nu}~,\nonumber\\
S\chi^\alpha_{q,s;\nu}&= 
\sum_{\alpha'=0}^{mn-1}  \sum_{\substack{\nu'\in\Z_{2(mn-1)}\\ \alpha'+\nu'=0\mod2}}\sum_{\substack{q',s'\\\alpha'+q'+s'=0\mod2}} S^{m,n}_{\alpha,\alpha'} e^{-2\pi i\frac{bb'}4}
e^{-2\pi i \left(- \frac{qq'}{2(mn+1)} + \frac{ss'}4 \right) }
\chi^{\alpha'}_{q',s';\nu'}~,
\end{align}
where 
we have introduced
\begin{align}
S^{m,n}_{\alpha,\alpha'}\equiv\frac1{2(mn+1)(mn-1)}  \sin \frac{\pi(\alpha+1)(\alpha'+1)}{mn+1}~.
\end{align}
Notice that the index $\nu$ enters the transformations \eqref{eq:LMtransfs} in a fairly trivial manner, that is, solely through the  
parafermion selection rule $\alpha+\nu=0\mod2$.
Thus, the modular transformations of the characters \eqref{eq:LMcharct}, at least formally, factor in four pieces. 
The index $\alpha$ transforms according to the characters of an affine $\su(2)$ 
system at level $mn-1$. 
The index $q$ transforms as a theta function at level $-(mn+1)$, while the index $s$ transforms as a theta function at level 2. 
These are very similar to the modular transformation properties of the
(2,2) minimal models \cite{Gepner:1987vz}.
The difference between our models and the minimal models is the 
extra term depending upon $b = s\mod2$, which 
exhibits the same transformation properties as a theta function at
level 2.  This is not an independent index of the characters,
but suggests the following way of thinking about these models.  A pair
of free left-moving fermions (realizing the $\SO(2)$ current algebra)
transforms precisely as a level 2 theta function (see
\eqref{eq:SO2ktransf} for $k=1$), and
exhibits the gravitational anomaly $c-\cb$ of our model.  As far as
modular properties are concerned, the $\GP_{m,n}$ model is equivalent
to an $A_{mn-1}$ minimal model together with a pair of left-moving
fermions, with the boundary conditions for the free fermions
tied to the boundary conditions for the
fermions in the minimal model (parameterized by $b$).  

This observation shows that we can actually consider a slightly more general partition function, when we take the indices $\alpha,\alphab$ to
be contracted with one of the known $\widehat{\su(2)}$ invariants at level
$mn-1$ \cite{Cappelli:1986hf}, 
obtaining the modular covariant partition function
\begin{align}
\label{eq:finalZGP}
Z&=\frac14 \sum_{\alpha=0}^{mn-1} \sum_{\nu\in\Z_{2(mn-1)}}\sum_{s\in\Z_4}\sum_{q\in \Z_{2(mn+1)}} 
N_{\alpha\alphab}^{(mn-1)}\chi^\alpha_{q,s;\nu} \chib^{\alphab}_{q,s}~,
\end{align}
where $N^{(k)}_{\alpha\alphab}$ is any $\widehat{\su(2)}$
invariant at level $k$.
The LG theory we have been considering thus far corresponds to the 
diagonal invariant $N^{(k)}_{\alpha\alphab}=\delta_{\alpha,\alphab}$, 
which exists at any level $k$.  More general modular covariant
combinations are possible; we provide a complete classification in the
next subsection.

We conclude this section by listing the expressions of the partition functions in the various sectors computed above in terms of our new notation
\begin{align}
\label{eq:GPpartfunctslist}
Z_{\text{AA}}&=\half \sum_{\alpha=0}^{mn-1} \sum_{\nu\in\Z_{2(mn-1)}}\sum_{q\in\Z_{2(mn+1)}}\sum_{s,\sb=0,2}
N_{\alpha\alphab}^{(mn-1)} \chi^\alpha_{q,s;\nu} \chib^{\alphab}_{q,\sb}~,\nonumber\\
Z_{\text{PA}}&=\half  \sum_{\alpha=0}^{mn-1} \sum_{\nu\in\Z_{2(mn-1)}}\sum_{q\in\Z_{2(mn+1)}}\sum_{s,\sb=0,2} e^{\pi i \left( s-\sb \right)}
N_{\alpha\alphab}^{(mn-1)}\chi^\alpha_{q,s;\nu}\chib^{\alphab}_{q,\sb}~,\nonumber\\
Z_{\text{AP}}&=\half \sum_{\alpha=0}^{mn-1} \sum_{\nu\in\Z_{2(mn-1)}}\sum_{q\in\Z_{2(mn+1)}}\sum_{s,\sb=\pm1}
N_{\alpha\alphab}^{(mn-1)}\chi^\alpha_{q,s;\nu} \chib^{\alphab}_{q,\sb}~,\nonumber\\
Z_{\text{PP}}&=\half \sum_{\alpha=0}^{mn-1} \sum_{\nu\in\Z_{2(mn-1)}}\sum_{q\in\Z_{2(mn+1)}}\sum_{s,\sb=\pm1}
e^{\pi i \frac{s-\sb}{2}}N_{\alpha\alphab}^{(mn-1)}\chi^\alpha_{q,s;\nu}\chib^{\alphab}_{q,\sb}~.
\end{align}
These transform under the modular group as
\begin{align}
\label{eq:TStransfSO2Z}
T \begin{pmatrix}
Z_{\text{AA}}	\\Z_{\text{PA}}		\\ Z_{\text{AP}}		\\ Z_{\text{PP}}
\end{pmatrix}
&=e^{-2\pi i\frac{1}{24}}\begin{pmatrix}
0	&1	&0					&0\\
1	&0	&0					&0\\
0	&0	&e^{\frac{\pi i }4}	&0\\
0	&0	&0					&e^{\frac{\pi i}4}
\end{pmatrix}
\begin{pmatrix}
Z_{\text{AA}}	\\Z_{\text{PA}}		\\ Z_{\text{AP}}		\\ Z_{\text{PP}}
\end{pmatrix}~,
&S \begin{pmatrix}
Z_{\text{AA}}	\\Z_{\text{PA}}		\\ Z_{\text{AP}}		\\ Z_{\text{PP}}
\end{pmatrix}&=
\begin{pmatrix}
1	&0	&0			&0\\
0	&0	&1			&0\\
0	&1	&0			&0\\
0	&0	&0			&-i
\end{pmatrix}
\begin{pmatrix}
Z_{\text{AA}}	\\Z_{\text{PA}}		\\ Z_{\text{AP}}		\\ Z_{\text{PP}}
\end{pmatrix}~.
\end{align}
As discussed above, these are precisely the transformation properties 
of a pair of left-moving free fermions with their spin structure
determined by $s\mod2$, as given in \eqref{eq:SO2ktilde} for $k=1$.

\subsection{Twisted boundary conditions and orbifolds}
\label{ss:orbifolds}

A particularly important role in our discussion will be played by the discrete symmetry of the theory. 
The transformations \eqref{eq:LMtransfs} show that the $\GP_{m,n}$ model admits a discrete symmetry group 
$G_{mn}\times\Gb_{mn}$, where $G_{mn}=\Z_{mn+1}\times\Z_2$
and the two factors act on the indices $q$ and $s$, respectively.
Explicitly, the fields of the theory are labeled by seven indices $\Phi^{\alpha,\alphab}_{q,s,\qb,\sb;\nu}$,\footnote{The
fields are determined in terms of a pairing between a collections of representations of the left-moving algebra and a (not necessarily complete)
representation of the right-moving $N=2$ algebra.} and
the action of $G_{mn}$ on these is given by
\begin{align}
\Phi^{\alpha,\alphab}_{q,s,\qb,\sb;\nu} &\rightarrow e^{2\pi i \frac{q}{mn+1}}\Phi^{\alpha,\alphab}_{q,s,\qb,\sb;\nu}~,
&\Phi^{\alpha,\alphab}_{q,s,\qb,\sb;\nu} &\rightarrow e^{2\pi i \frac{s}{2}}\Phi^{\alpha,\alphab}_{q,s,\qb,\sb;\nu}~.
\end{align}
In particular, only the diagonal subgroup of $G_{mn}\times\Gb_{mn}$ acts non-trivially on \eqref{eq:finalZGP}. 
It is therefore possible to 
generate new modular covariant theories
by taking quotients of the model by any subgroup of the diagonal
$G_{mn}\subset G_{mn}\times \Gb_{mn}$ (or equivalently $\Gb_{mn}$). In
fact, this provides a
complete classification of all modular covariants, because of the
known properties of theta functions
\cite{Gepner:1986hr}.  Further, as shown for minimal models in those
references and applied in \cite{Greene:1990ud}, a quotient by the
complete group $G_{mn}$ produces an isomorphic theory, obtained by
reversing the signs of $\qb$ and $\sb$, or equivalently simply
reversing the sign of $\Jb_0$.  For the interested reader, we
explicitly demonstrate this below.

In order to do that, it is convenient to collect both indices $q$ and $s$ into a single vector $\bmu=(q,s)$ equipped with the product
\begin{align}
\bmu\cdot\bmu' = -\frac{\mu_1\mu_1'}{2(mn+1)}+\frac{\mu_2\mu_2'}4~.
\end{align}
Now, let $\bgamma$ be the generator of a subgroup $\Z_h\subseteq G_{mn}$, then we define
the twisted partition function
\begin{align}
\label{eq:Ztwist}
Z[x,y]&=\frac14 \sum_{\alpha,\alphab}\sum_{\nu\in\Z_{2(mn-1)}}\sum_{\bmu}
N_{\alpha\alphab}^{(mn-1)}e^{-4\pi ix \bgamma\cdot(\bmu+\bgamma y) } \chi^\alpha_{\bmu;\nu}
\chib^{\alphab}_{\bmu+2y\bgamma }~.
\end{align}
where $x,y\in\Z_h$.
It is possible to show that this behaves covariantly under modular transformations
\begin{align}
\label{eq:twisttrasprop}
TZ[x,y]&=e^{-2\pi i\frac1{24}}\left(Z_{\text{AA}}[x+y,y]+Z_{\text{PA}}[x+y,y]+e^{2\pi i\frac18}\left(Z_{\text{AP}}[x+y,y]+Z_{\text{PP}}[x+y,y]\right)\right)~,\nonumber\\
SZ[x,y]&=Z_{\text{AA}}[y,x]+Z_{\text{PA}}[y,x]+Z_{\text{AP}}[y,x]+e^{-2\pi i\frac14}Z_{\text{PP}}[y,x]~,
\end{align}
that is, $Z[x,y]$ satisfies \eqref{eq:TStransfSO2Z}, where the twisted partitions functions over the various sectors are constructed from \eqref{eq:Ztwist}
by restricting the indices as in \eqref{eq:GPpartfunctslist}.
Next, we construct the orbifold theory by summing over the various twisted sectors
\begin{align}
\label{eq:Zhorb}
Z_{h}&=\frac{1}{h} \sum_{x,y\in\Z_{h}} Z[x,y]\nonumber\\
&=\frac{1}{4h} \sum_{x,y\in\Z_{h}} \sum_{\alpha,\alphab}\sum_{\nu\in\Z_{2(mn-1)}}\sum_{\bmu}
N_{\alpha\alphab}^{(mn-1)}e^{-4\pi ix \bgamma\cdot(\bmu+\bgamma y) } \chi^\alpha_{\bmu;\nu}
\chib^{\alphab}_{\bmu+2y\bgamma}~.
\end{align}
The properties \eqref{eq:twisttrasprop} ensure that the orbifold partition function exhibits the same modular covariant 
transformation properties \eqref{eq:TStransfSO2Z}
as the original untwisted theory.

As mentioned above, a particularly relevant case is realized by taking the quotient by the full symmetry group $G_{mn}$.
We can achieve this in two steps. First, we implement the orbifold by the subgroup $\Z_{mn+1}\subset G_{mn}$, which 
is generated by $\bgamma=(1,0)$. In this case \eqref{eq:Zhorb} reads
\begin{align} 
Z_{mn+1}&=\frac{1}{4(mn+1)} \sum_{x,y\in\Z_{mn+1}} \sum_{\alpha,\alphab}\sum_{\nu\in\Z_{2(mn-1)}}\sum_{\bmu}
N_{\alpha\alphab}^{(mn-1)}e^{-2\pi ix \frac{-2q-2y}{2(mn+1)}} \chi^\alpha_{q,s;\nu}
\chib^{\alphab}_{q+2y,s}~.
\end{align}
We can carry out the sum over $x$ independently
\begin{align}
\sum_{x\in\Z_{mn+1}}e^{-2\pi ix \frac{-2q-2y}{2(mn+1)}}=
\begin{cases}  mn+1 & \text{if }  2q+2y\in 2(mn+1)\Z  \\
0 &\text{otherwise} 
\end{cases}~,
\end{align}
which yields the condition $2y=-2q\mod 2(mn+1)$. Plugging this in we obtain 
\begin{align} 
Z_{mn+1}&=\frac{1}{4} \sum_{\alpha,\alphab}\sum_{\nu\in\Z_{2(mn-1)}}\sum_{\bmu}
N_{\alpha\alphab}^{(mn-1)} \chi^\alpha_{q,s;\nu}
\chib^{\alphab}_{-q,s}~,
\end{align}
which is simply the partition function we started from up to reversing the charge $\qb$ for all states. 
In order to perform the quotient by the full $G_{mn}$ we need to further quotient the theory we just obtained by the remaining $\Z_2$ factor.
The procedure is practically identical to the one we just performed, thus we skip the derivation, and we instead simply present the final result
\begin{align} 
Z_{G_{mn}}&=\frac{1}{4}
\sum_{\alpha,\alphab}\sum_{\nu\in\Z_{2(mn-1)}}\sum_{\bmu}
N_{\alpha\alphab}^{(mn-1)} \chi^\alpha_{q,s;\nu}
\chib^{\alphab}_{-q,-s}~.
\end{align}
This describes exactly the untwisted theory we started from, with the signs of both
$\qb$ and $\sb$ for all states reversed.
Hence, this partition function is related to the original one
by the operation $\bqb\rightarrow-\bqb$ for all states. 
Thus, $Z$ and $Z_{G_{mn}}$ define isomorphic conformal field theories,
the isomorphism being realized by reversing the sign of all $\GU(1)_R$ charges. 
This proves that, as in the $N=2$ minimal models case, the order-disorder duality 
of the parafermionic theory \cite{Gepner:1986hr} extends to the class of $\GP_{m,n}$ models.

\section{Exactly solvable heterotic string theories}
\label{s:exactstrings}

In this section we tackle the problem of employing the models studied in detail above
to construct spacetime supersymmetric heterotic compactifications. 
In essence, our observation that the characters transform in the same
way as those of a minimal model and a pair of left-moving free
fermions with their spin structure tied allows us to use previous
results \cite{Gepner:1987vz,Lutken:1988hc}.  The celebrated
construction of Gepner begins with a tensor product of $R$ minimal models
such that the total central charge is $c=\cb=9$.  This is then
combined with additional 10 free fermions realizing
$\widehat{\so(10)}$.  The spin structures of the fermions and of all 
$R$ minimal models are tied by a collection of suitable $\Z_2$
projections, and a further projection to integral $\bqb$ in the NS
sector produces a modular covariant theory with $(c,\cb)=(14,9)$ which,
when tensored with the hidden $\widehat{\mathfrak{e}_8}$ current algebra yields the
internal theory for a heterotic compactification.  This is then
combined with the free fields describing spacetime with a twist
due to \cite{Gepner:1987qi} relating superstring vacua of this type to
their heterotic versions, and GSO projected
to produce a supersymmetric heterotic string compactification to four
dimensions with unbroken gauge symmetry $\GE_6\times \GE_8$.

What we have seen is that the characters of the generalized $\GP_{m,n}$
model (with arbitrary $\widehat{\su(2)}$ invariant at level $k=mn-1$)
transform precisely the same way as those of the associated minimal
model at level $k$ tensored with two free fermions, with the spin
structures tied together.  This means that starting with any exactly
solved Gepner model we can construct new (0,2) models by replacing $N_2$
of the minimal model components, so long as their levels are such that
$k_j+1$ is not prime, along with $N_2$ pairs of left-moving fermions, by
$N_2$ generalized $\GP_{m_j,n_j}$ models, $j=1,\dots,N_2$, such that $m_j n_j = k_j+1$ (in
general there may be more than one choice of $m,n$ for a given $k$ and these
will lead to distinct models).  Repeating the Gepner construction
described above will produce a supersymmetic heterotic
compactification to four dimensions with an unbroken gauge group of
rank $6-N_2$ which contains $\SO(10-2N_2)\times \GU(1)_L$ as a maximal
subgroup.  These are often denoted $\GE_{6-N_2}$.\footnote{In this notation 
$\GE_5=\SO(10)$, $\GE_4=\SU(5)$, $\GE_3=\SU(3)\times\SU(2)$.}  This
leads to a large collection of exactly solved heterotic vacua in four
dimensions; with minor modifications the same applies to Gepner models
leading to six- and to eight-dimensional theories.  Below we provide
the explicit construction for the interested reader.

We consider the tensor product of $N_1$ $N=2$ minimal models and $N_2=R-N_1$ generalized GP${}_{m,n}$ models, such that 
\begin{align}
\label{eq:cbar}
\cb = \sum_{a=1}^{R}\frac{3k_a}{k_a+2}=3d~,
\end{align}
where we have defined $k_{j+N_1} = m_j n_j-1$.
Since we are interested in constructing (0,2) models, we assume that $N_2\geq1$.
We are going to show that this theory, for which $c=3d+N_2$,
can be completed to a consistent heterotic compactification to the extended spacetime $\R^{1,9-2d}$ 
with spacetime gauge group $\GE_{9-d-N_2}\times\GE_8$.

The first step in our strategy is to construct a modular invariant partition function. 
Then, we will attain a consistent heterotic theory by performing a series of quotients to impose
spacetime supersymmetry while preserving modular invariance. 
Let us implement explicitly the above line of reasoning.
We start with the product theory
\begin{align}
\label{eq:Zprod}
Z_{\text{prod}}&=\frac14\sum_{\balpha,\balphab}\sum_{\bmu} N_{\balpha\balphab}\chi^{\balpha}_{\bmu;\bnu} \chib^{\balphab}_{\bmu+\bmu_0}~.
\end{align}
where we collected the various indices of the component theories into the vectors $\balpha=(\alpha_1,\dots,\alpha_R)$, $\bmu=(\mu_0,q_1,\dots,q_R,s_1,\dots,s_R)$
and $\bnu=(\nu_1,\dots,\nu_{N_2})$, and the characters are given by
\begin{align}
\label{eq:prodchars}
\chib^{\balphab}_{\bmub}&\equiv B^{(8-2d)}_{\mub_0} \left(\prod_{a=1}^{R} \chib^{\alphab_a}_{\mub_a,\mub_{a+R}}\right)
~,\nonumber\\
\chi^{\balpha}_{\bmu;\bnu}&\equiv \chi^{\GE_8}_0 B^{(16-2d-2N_2)}_{\mu_0} \left(\prod_{i=1}^{N_1} \chi^{\alpha_i}_{\mu_i,\mu_{i+R}}\right)
\left(\prod_{j=N_1+1}^{R} \chi^{\alpha_j}_{\mu_j,\mu_{j+R};\nu_j}\right)~,
\end{align}
where $\chi^{\GE_8}_0$ is the character for the singlet
$\widehat{\mathfrak{e}_8}$ representation; $B^{(2k)}_\lambda$ are the $\SO(2k)$
characters (see \eqref{eq:Blambdas}) with the association
$(0,1,2,3)\rightarrow(\text{o},\text{s},\text{v},\overline{\text{s}})$; and
\begin{align}
N_{\balpha\balphab} = \prod_{a=1}^{R} N^{(k_a)}_{\alpha_a\alphab_a}
\end{align}
is the product of the $\widehat{\su(2)}$ invariants of the component
theories.

The index $a=0$ on the left-movers refers to the $\SO(16-2d-2N_2)$
representation.\footnote{As
noted in the appendix, the various $\widehat{\so(2k)}$ algebras we
encounter will all be realized at level one.  We will thus be
slightly careless and speak of $\SO(2k)$ representations when there
is no risk of confusion.}  On the right it refers to the $\SO(8-2d)$ representation
carried by the light-cone gauge spacetime fermions.
The vector $\bmu_0=(2,0,0,\dots)$ implements the permutation of the
representations of this that implements the result of
\cite{Gepner:1987qi}.\footnote{In Gepner's construction, this was used
to convert a superstring compactification to a heterotic one.  Our
models do not admit a superstring interpretation.  We follow this
procedure to obtain a modular invariant partition function based on the relation
mentioned at the start of this section between our models and the
heterotic models constructed from $N=2$ minimal models.}

We also introduce the product
\begin{align}
\label{eq:LMchprod}
\bmu\cdot\bmu'&=\frac{\mu_0\mu_0'}4+\sum_{a=1}^{R}\left( -\frac{\mu_a\mu_a'}{2(k_a+2)}+\frac{\mu_{a+R}\mu_{a+R}'}4\right)~.
\end{align}
In terms of this the expressions for the transformation properties under the modular group of the product characters
introduced in \eqref{eq:prodchars} take a particularly compact form.
A combination of \eqref{eq:LMtransfs} and \eqref{eq:SO2ktransf} results in
\begin{align}
\label{eq:LMtransftot}
T\chi^{\balpha}_{\bmu;\bnu}&=e^{-2\pi i \frac{16+2d}{24}} e^{\pi
  i\bmu\cdot\bmu} T_\balpha e^{\pi i\left((7-d-N_2) \frac{b_0^2}4+\sum_{a=N_1+1}^R\frac{b_{a}^2}4 \right)}\chi^{\balpha}_{\bmu;\bnu}~,\nonumber\\
S\chi^{\balpha}_{\bmu;\bnu}&= \sum_{\balpha',\bmu',\bnu'} S_{\balpha,\balpha'} e^{-2\pi i \bmu\cdot\bmu'}
e^{-2\pi i\left((7-d-N_2) \frac{b_0 b_0'}4+\sum_{a=N_1+1}^R\frac{b_{a}b_{a}'}4 \right)} \chi^{\balpha'}_{\bmu';\bnu'}~,
\end{align}
where $b_0\equiv\mu_0\mod2$ and $b_a\equiv\mu_{a+R}\mod2$, $a=1,\dots,R$, and 
\begin{align}
T_{\balpha}&= \prod_{a=1}^{R} e^{2\pi i \frac{\alpha_a(\alpha_a+2)}{4(k_a+2)}}~,\nonumber\\
S_{\balpha,\balpha'}&=\half\prod_{a=1}^{R}\frac1{2k_a(k_a+2)}  \sin \frac{\pi(\alpha_a+1)(\alpha_a'+1)}{k_a+2}~.
\end{align}
The modular transformations of the right-moving characters in this
notation read
\begin{align}
\label{eq:TStransfRMprod}
T\chib^{\balphab}_{\bmub}&=e^{2\pi i \frac{4+2d}{24}} \Tb_{\balphab} e^{-\pi i\bmub\cdot \bmub}e^{-\pi i(3-d) \frac{b_0^2}4}\chi^{\balphab}_{\bmub}~,\nonumber\\
S\chib^{\balphab}_{\bmub}&= \sum_{\balphab',\bmub'} S_{\balphab,\balphab'} e^{2\pi i \bmu\cdot\bmu'}
e^{2\pi i(3-d) \frac{b_0 b_0'}4} \chi^{\balphab'}_{\bmub'}~.
\end{align}

Unfortunately, the product theory defined by \eqref{eq:Zprod} does not lead to a modular invariant partition function, due to the covariance
of the transformations \eqref{eq:TStransfSO2Z} 
of the GP${}_{m,n}$ models. 
In fact, as we spelled out in the previous section, a single $\GP_{m,n}$ model transforms under the modular group as 
certain linear combinations of $\SO(2)$ representations. It is possible to achieve modular invariance 
by coupling the spin structures of the  $\GP_{m,n}$ component theories and of the left-moving $\SO(16-2d-2N_2)$ representations.
Upon introducing twisted sectors as required by modular invariance, we arrive at the partition function
\begin{align}
\label{eq:hetpfmodinv}
Z_{\text{proj}}=\frac14 \sum_{\substack{\balpha,\balphab,\bmu,\bmub,\bnu\\ \bmu-\bmub\in\Lambda_0\\2\bgamma_a\cdot\bmub\in\Z}} N_{\balpha\balphab}e^{\pi i b_0} \chi^{\balphab}_{\bmu;\bnu}\chib^{\balphab}_{\bmub+\bmu_0}~,
\end{align} 
where $\Lambda_0$ is the lattice generated over $\Z$ by
$2\bgamma_a\equiv2\bdelta_{0}+2\bdelta_{R+a}$, $a=N_1+1,\dots,R$, where $(\bdelta_i)_j=\delta_{ij}$. 
We show explicitly that $Z_{\text{proj}}$ is indeed modular invariant
in appendix \ref{app:modinv}.

We can now use the method introduced in \cite{Gepner:1987qi} to
construct quotients of \eqref{eq:hetpfmodinv} by subgroups of the
discrete symmetry $G_{\text{proj}}\equiv \left(\prod_{i=1}^{N_1}
  G_{k_i+1}\right)\times\left(\prod_{j=1}^{N_2}
  \Z_{m_jn_j+1}\right)$.  Note that the $\Z_2\subset G_{m_jn_k+1}$
have been removed by our construction.  
Let $\Z_h$ be a subgroup of $G_{\text{proj}}$ generated by $\bgamma$.
Then we define the twisted partition function
\begin{align}
\label{eq:Zprodtwist}
Z_{\text{proj}}[x,y]&=\frac14\sum_{\balpha,\balphab}\sum_{\substack{\bmu,\bmub,\bnu\\ \bmu-\bmub\in\Lambda_0\\2\bgamma_a\cdot\bmub\in\Z}} N_{\balpha\balphab}e^{\pi i b_0} e^{-4\pi ix \bgamma\cdot(\bmu+\bgamma y) }\chi^{\balpha}_{\bmu;\bnu} \chib^{\balphab}_{\bmu+\bmu_0+2y\bgamma}~.
\end{align}
This exhibits the correct modular properties, that is, $TZ_{\text{proj}}[x,y]=Z_{\text{proj}}[x+y,y]$ and 
$SZ_{\text{proj}}[x,y]=Z_{\text{proj}}[y,x]$, and it depends on $x,y$
only mod $h$.  As shown in \cite{Greene:1990ud} for suitable $\bgamma$
this leads to consistent projections in twisted sectors, and can be
repeatedly applied for a collection of mutually consistent projections.

Now, as part of the procedure to obtain a consistent string vacuum, compatibility with the 
superconformal gauge condition requires that the
spin structures for all the component theories are tied together \cite{Mueller:1986yr,Gepner:1987qi}. 
This is achieved by a series of $\Z_2$ quotients, which in our formalism 
are represented by the vectors $\bgamma_i=\bdelta_{0}+\bdelta_{R+i}$, $i=1,\dots,N_1$.
The resulting theory is restricted to $\bmu,y$ such that
 \begin{align}
 2\bgamma_a\cdot \left(\bmu+y\bgamma_a\right)= \frac{\mu_0+s_a}2+y\in\Z~,	\qquad\qquad  a=1,\dots,R~.
 \end{align}
Since by construction $y\in\Z$, this implies that $\mu_0=s_a\mod 2$.

Next, we need to project out states which have non-integral charges under the $\GU(1)_L\times\GU(1)_R$ symmetry. This ensures that the 
right-moving $N=2$ superconformal algebra is unbroken, and that the linearly realized $\GU(1)_L\times\SO(16-2d-2N_2)$ symmetry enhances to the
full $\GE_{9-d-N_2}\times\GE_8$ spacetime gauge group.
Since in our original partition function \eqref{eq:hetpfmodinv} the
charges satisfy $\bq-\bqb\in\Z$, it suffices to quotient by $\bgamma_0=(1,\dots,1)$, 
corresponding to the orbifold of the theory by the operator $e^{2\pi i J_0}$ (or equivalently $e^{2\pi i \Jb_0}$).
In fact, this implies
\begin{align}
\label{eq:Jb0proj}
2\bgamma_0\cdot \left(\bmu+ y\bgamma_0\right)= \bqb+2y\in\Z~,
\end{align}
that is, $\bqb\in\Z$ as desired.
The final step involves a GSO projection to definite fermion number. A suitable choice for a fermion number operator, due to the projection to 
integral charges in the previous step, is the operator $e^{\pi i \Jb_0}$. In particular, we wish to project onto negative total $\bqb$ charge, that is, $e^{\pi i \Jb_0}=-1$. 
Finally, this leads to the partition function of our heterotic model
\begin{align}
\label{eq:hetpartfunct}
Z=\frac14 \sum_{\substack{\balpha,\balphab,\bmu,\bmub,\bnu\\ \bmu-\bmub\in\Lambda}} e^{\pi i b_0} 
N_{\balpha\balphab} \chi^{\balphab}_{\bmu;\bnu}\chib^{\balphab}_{\bmub+\bmu_0}~,
\end{align} 
where the sum is over $\bmu$ such that $2\bgamma_0\cdot\bmu\in2\Z+1$ and $2\bgamma_a\cdot\bmu\in\Z$, $a=1,\dots,R$, 
and where $\Lambda$ is the lattice generated over $\Z$ by
$\bgamma_0, 2\bgamma_a$.
We refer to the theories defined by \eqref{eq:hetpartfunct} as GP models.

\subsection{The mirror duality}

We have already highlighted that the form of the partition function 
\eqref{eq:hetpartfunct} is strongly reminiscent of
that of Gepner models \cite{Gepner:1987qi}. 
Formally, in fact, the only difference is the appearance of the index $\bnu$ in the left-moving characters. 
This index, however, did not participate in any of the steps of our procedure to derive the partition function \eqref{eq:hetpartfunct}.
This is due to the fact that the discrete symmetry group $G_{\text{proj}}$ acts on the $\bmu$ indices alone, 
and $\bnu$ does not make any relevant contribution to the modular transformations \eqref{eq:LMtransftot}.
In particular, the subgroup of $G_{\text{proj}}$ which is preserved by our final theory is $G\times\Z_2$, where
\begin{align}
G&=\frac{\left(\prod_{i=1}^{N_1} (\Z_{k_i+2})\right)\times 
\left(\prod_{j=1}^{N_2} (\Z_{m_jn_j+1})\right)}{\Z_{p}}~,
\end{align}
and equivalently for the right-moving sector of the theory.
The quotient by $\Z_p$, where $p$ is the order of the cyclic group
generated by $e^{2\pi i (J_0+\Jb_0)}$, is due to our projection \eqref{eq:Jb0proj},
while the $\Z_2$ factor is simply charge conjugation.

These observations, together with the order-disorder duality of $\GP_{m,n}$ models we proved in section \ref{ss:orbifolds},
 imply that formally our theories share all the properties of Gepner models which lead to the Greene-Plesser mirror construction. 
Hence, it appears natural to conjecture that the set of orbifolds of a GP model is organized into pairs of mirror theories, which are related by
right-moving charge conjugation. 
In the remainder of this section we will briefly sketch the original argument 
of \cite{Greene:1990ud} applied to the theories under study, showing that this is indeed the case.

Starting with the theory defined by \eqref{eq:hetpartfunct}, it is possible to construct new modular invariant conformal field theories 
by taking appropriate quotients by discrete symmetries of the theory \cite{Gepner:1987qi}. 
One particular instance of this is given by the quotient by the full $G$.
It is not hard to see that the resulting partition function $Z_G$ will involve only
states satisfying $\bmub=-\bmu$. This new partition function, however, will not be modular invariant. 
Applying to $Z_G$ the procedure of the previous section we will obtain a modular invariant partition function
in which all states satisfy $\bmub=-\bmu$. This theory is of course not new, since it is related to our original partition function
by right-moving charge conjugation, thus defining an isomorphic theory. 
Finally, we recall that elements 
$(\gamma_1,\dots,\gamma_{R})\in G$ that do not satisfy the condition
\begin{align}
\label{eq:mirquotcond}
\sum_{i=1}^{N_1} \frac{\gamma_i}{k_i+2}+\sum_{j=1}^{N_2} \frac{\gamma_{N_1+j}}{m_jn_j+1}\in\Z
\end{align}
do not survive the projection to integral charges.  Then, let $H$ denote the subgroup of $G$
whose elements satisfy the condition \eqref{eq:mirquotcond}.  
The quotient by $G$ followed by the projections described will produce
in all a quotient by $H$.
The argument above leads to the conclusion that $Z$ and $Z_H$
are isomorphic theories, the isomorphism being right-moving charge conjugation, hence constituting
a mirror pair. This assertion extends unaltered to the whole set of orbifolds of $Z$,\footnote{This set comprises
orbifolds by symmetries not contained in $G$ as well.} which is therefore organized in pairs of mirror dual theories, as claimed.

\subsection{DK models}

In this section we turn to the geometric interpretation of the theories we constructed. 
This is realized in terms of a (0,2) NLSM on $\cE\rightarrow M$, 
where $\cE$ is a rank-$(d+N_2)$ holomorphic vector bundle
over the (possibly singular) 
complete intersection Calabi-Yau $d$-fold $M$.
Specifically, we are going to show that the GP model obtained as the $\Z_{d_P}$ orbifold of
\begin{align}
\label{eq:GPmodDK}
A_{k_1}\oplus\cdots\oplus A_{k_{N_1}}\oplus\GP_{m_1,n_1}\oplus\cdots\oplus\GP_{m_{N_2},n_{N_2}}~,
\end{align}
where $d_P$ is the least common multiple of $k_i+2,m_jn_j+1$, and where we take the diagonal $\widehat{\su(2)}$ affine invariant for all the component theories, 
corresponds to a codimension $2N_2$ CICY
of degrees $m_jn_jb_j,(m_j+n_j)b_j$, $j=1,\dots,N_2$, in a weighted projective space. 

Our strategy follows the procedure of \cite{Distler:1993mk}, namely we construct a $\GU(1)$ linear model\footnote{Our conventions
for (0,2) linear models follow \cite{Bertolini:2014dna}.} realizing
a LGO phase which, for an appropriate choice of the (0,2) superpotential, reduces to the GP model \eqref{eq:GPmodDK}. 
The other phase will be described in terms of a (0,2) NLSM with target space the complete intersection above.  

Explicitly, let us introduce $N_1+2N_2+1$ bosonic (0,2) chiral supermultiplets $X^i, Y^a_j,P$, where $i=1,\dots,N_1$, $j=1,\dots,N_2$ and $a=1,2$, 
as well as $N_2-1$ additional auxiliary (0,2) chiral supermultiplets $Z_\mu$, $\mu=1,\dots,N_2-1$.
We couple these to a $\GU(1)$ gauge group through the following set of charges
\begin{align}
\label{eq:DKGLSMbos}
\xymatrix@R=1mm@C=6mm{
\text{fields}	&X_i		&Y_{1j}	&Y_{2j}	&Z_\mu	&P		&\text{F.I.}\\
\GU(1)		&a_i		&m_jb_j	&n_jb_j	&d_P	&-d_P	&r
}
\end{align}
where we defined
\begin{align}
a_i&\equiv \frac{d_P}{k_i+2}~,\qquad i=1,\dots,N_1~,	&b_j&\equiv \frac{d_P}{m_jn_j+1}~,\qquad  j=1,\dots,N_2~.
\end{align}
We also introduce (0,2) chiral Fermi superfields $\Gamma^i,\Gammat^j,\Gammah^\mu,\Lambda_j^a$ with gauge charges
\begin{align}
\xymatrix@R=1mm@C=6mm{
\text{fields}	&\Gamma^i		&\Gammat^j	 	&\Gammah^\mu	&\Lambda_j^1		&\Lambda_j^2\\
\GU(1)		&a_i				&-m_jn_jb_j+d_P	&0	&-m_jn_jb_j		&-(m_j+n_j)b_j
}
\end{align}
These fields interact through the superpotential 
\begin{align}
\cL_{\text{sup}}&=\int d\theta \left[ \sum_{i=1}^{N_1} \Gamma^i P J_i(X,Y) +\sum_{\mu=1}^{N_2-1}\Gammah^\mu P \Jh_\mu 
+\sum_{j=1}^{N_2}\left( \Gammat^j P \Jt_j + \sum_{a=1}^2 \Lambda_j^a H^j_a\right)  \right]+\text{h.c.}~,
\end{align}
where the various functions we introduced are homogenous polynomials in the fields $X,Y,Z$ of degrees given by
\begin{align}
\xymatrix@R=1mm@C=6mm{
\text{sup.}		&J_i				&\Jt_j	 		&\Jh_\mu			&H^j_1			&H^j_2\\
\text{degree}	&d_P-a_i			&m_jn_jb_j		&d_P			&m_jn_jb_j		&(m_j+n_j)b_j
}
\end{align}
The model admits the following chiral symmetries
\begin{align}
\xymatrix@R=1mm@C=8mm{
\text{fields}	&X^i		&Y^a	&P		&\Gamma^I		&\Lambda^A		&\theta\\
\GU(1)_L		&0		&0		&1		&-1				&0				&0\\
\GU(1)_R		&0		&0		&1		&0				&1				&1
}
\end{align}
It is easy to check that the quadratic gauge anomaly vanishes, while the mixed $\GU(1)_G-\GU(1)_{L,R}$ anomalies also vanish
when the model satisfies the additional condition
\begin{align}
\sum_{i=1}^{N_1}\frac{1}{k_i+2}+\sum_{j=1}^{N_2}\frac1{m_jn_j+1}=1~.
\end{align}
The phase structure of this model is very simple. At $r\ll0$ the D-terms force $p$, the lowest component of the superfield $P$, to acquire a non-zero vev, which breaks
the gauge group down to the invariant $\Z_{d_P}$ subgroup. The F-terms force all the other fields to vanish,
and we recover the Landau-Ginzburg orbifold theory which realizes the GP model \eqref{eq:GPmodDK} for the specific form of the superpotential
\begin{align}
\label{eq:GPsupform}
J_i&=x_i^{k_i+1}~,	&\Jh_\mu&=z_\mu~,		&\Jt_j&=y_{1j}^{n_j}~,	&H^j_1&=y_{2j}^{m_j}~,		&H^j_2&=y_{1j}y_{2j}~.
\end{align}
In fact, $(Z_\mu,\Gammah^\mu)$, when present, constitute pairs of massive fields and have no effect on the IR theory, whose non-trivial dynamics is then
described by the GP model \eqref{eq:GPmodDK}.

In the $r\gg0$ phase, instead, the fields $x,y,z$ are forced not to simultaneously vanish, and upon taking the quotient by the $\GU(1)$ gauge group, 
these parametrize the $(N_1+3N_2-2)$-dimensional
weighted projective space $V=\P^{N_1+3N_2-2}_{\{a_i\},\{m_jb_j,n_jb_j\},\{d_P\}}$ whose weights are proportional 
to the gauge charges \eqref{eq:DKGLSMbos}.
The F-terms instead force $<p>=0$, and for sufficiently generic 
$H^j_a$, the space of classical vacua of the theory is the (possibly singular)
complete intersection $M=\{ (x,y,z)\in V | H^j_a(x,y,z)=0\}$.
If $M$ is non-singular, the fermions $\lambda^j_a$ -- the lowest components of $\Lambda^j_a$ -- all acquire a mass 
and the massless left-moving fermions transform as sections of the rank-$(d+N_2)$ bundle $\cE\rightarrow M$ defined by the SES
\begin{align}
\label{eq:LMbundle}
\xymatrix@R=0mm@C=12mm{
0	\ar[r]	&\cE	\ar[r]	&{\begin{matrix} \oplus_{i=1}^{N_1}\cO(a_i)\\\oplus\\ \oplus_{j=1}^{N_2}\cO(-m_jn_jb_j+d_P)\\\oplus\\\cO^{\oplus (N_2-1)}\end{matrix}} \ar[r]^-{J}	&\cO(d_P)	  \ar[r]	&0~.
}
\end{align}
restricted to $M$. 

This construction will not always yield
a smooth $M$. There are two potential sources of singularities. 
The CICY might intersect some of the ambient weighted projective variety orbifold singularities, if these appear in codimension low enough.
This type of singularities is well-understood in (2,2) theories \cite{Aspinwall:2010ve}, and partially under control in the (0,2) context as well \cite{Distler:1996tj}. 
A second possibility is that the complete intersection exhibits complex structure 
singularities even for generic values of the defining equations. 
This happens when the monomials in $H^j_a$ are not allowed by gauge invariance to be generic enough 
in order for the complete intersection to be transverse.
A novel feature of (0,2) model is that in some cases 
the classical space of vacua remains
nonetheless compact and it is expected that the corresponding CFT is non-singular \cite{Chiang:1997kt}. 

As it turns out, from the point of view of our mirror duality we are forced to study such singular loci, even when $M$ is generically non-singular.
In fact, setting the (0,2) superpotential to its GP form \eqref{eq:GPsupform} the resulting CICY defined by $H^j_1=H^j_2=0$ 
develops complex structure singularities along the locus $y_{1j}=y_{2j}=0$. On the other hand,
deforming the superpotential to obtain a non-singular complete intersection always breaks the symmetry group $H$. 
Thus, the quotient that yields the mirror dual theory does not have a
geometric interpretation in terms of an orbifold of a smooth compact CY.
The locus in moduli space preserved by $H$ is associated to singular CICYs. In the next section we will study
this locus in detail, provide an interpretation for it
and thus show that our mirror quotient does in fact have a geometric realization.

\subsection{Singular CICYs and hybrid models}
\label{ss:hybrids}

A key property of (2,2) mirror symmetry is that the duality obtained by quotienting the theory at the Gepner point
naturally extends to geometry. In fact, whenever a geometry interpretation exists \cite{Aspinwall:1994cf},
the symmetry group $H$ of the Gepner model\footnote{We do not introduce new notation here since Gepner models can be considered
a special case of our construction, realized by taking $N_2=0$.} upon which the mirror procedure is based comprises exactly those
symmetries which turn out to be geometrical. 
As we discussed in the previous section, for our (0,2) models it is not possible to describe a transverse complete intersection
while preserving $H$ as a symmetry. 
To seek a geometric interpretation for our quotient construction, it is natural to begin with
a closer look at the $H$-preserving locus in the geometric $r\gg 0$ phase of the linear model. 

We start by noting that upon setting the superpotential to its GP form \eqref{eq:GPsupform} in the $r\gg0$ phase, a combination of 
D- and F-term effects forces $<p>=0$, implying that the classical space of vacua remains compact.  
Thus, we find ourselves in the possibility referred to in the previous section: 
despite describing a singular complete intersection, the GP locus does not lead to a decompactification in the linear model.
However, the question remains: does this locus correspond to a
non-singular CFT and, if so, do we have a useful description of it?
In the remainder of this section we will show that the answer to both questions is yes.

The second observation is that every element of $H$ acts homogeneously on the monomials $y_{1j}^{n_j}$ and $y_{2j}^{m_j}$. 
This follows from the fact that the $\Z_{mn+1}\subset G$ symmetry of the GP${}_{m,n}$ model acts as
\begin{align}
y_1&\rightarrow \xi^{m} y_1~,		&y_2&\rightarrow \xi^{n} y_1~,
\end{align} 
where $\xi=e^{\frac{2\pi i}{mn+1}}$. Thus, the $H$-preserving locus admits, in particular, the following class of deformations of the superpotential
\begin{align}
\label{eq:02hybridsup}
H_1^j\rightarrow H_1^j{}'&=a_jy_{1j}^{nj}+y_{2j}^{m_j}~,		&H_2^j&=y_{1j}y_{2j}~,
\end{align}
for any set of coefficients $a_j$.

Now, if we assume  $a_j\neq0$, imposing $H_1^j{}'=H_2^j=0$ implies $y_{1j}=y_{2j}=0$. 
Upon quotienting by the $\GU(1)$ gauge group
the $x,z$ coordinates describe a compact manifold $B=\P^{N_1+N_2-2}_{\{a_i\}\{d_P\}}$, while the massless fields $p,y_{1j},y_{2j}$
transform as appropriate line bundles over $B$. This is precisely the structure of a hybrid model \cite{Bertolini:2017lcz} $\cE\rightarrow\bY$,
where the target space is given by
\begin{align}
\label{eq:Yhybdef}
\bY=\tot\left(\cO(-d_p)\oplus\oplus_{j=1}^{N_2}\left(\cO(m_jb_j)\oplus\cO(n_jb_j)\right)\rightarrow B \right)~,
\end{align}
equipped with a bundle determined by \eqref{eq:LMbundle}, which in particular splits to a sum of line bundles
\begin{align}
\label{eq:Ehybdef}
 \cE = \oplus_{i=1}^{N_1}\cO(a_i)\oplus\cO^{\oplus(N_2-1)} \oplus_{j=1}^{N_2}\left( \cO(-m_jn_jb_j+d_P)\oplus\cO(-m_jn_jb_j)\oplus\cO((m_j+n_j)b_j)\right)~.
\end{align}
More precisely,
the various summands in \eqref{eq:Ehybdef} are line bundles over $\bY$ described in terms of pullbacks of line bundles over $B$, 
but we omit the pullback map to simplify the notation.
The hybrid model moreover inherits from the GLSM a non-trivial superpotential
\begin{align}
\bJ\equiv \begin{pmatrix}
pJ_i 	 &p\Jh_\mu		&p\Jt_j	&H_1^j{}'	&H_2^j
\end{pmatrix} \in \Gamma(\cE^\ast)
\end{align}
which satisfies $\bJ^{-1}(0)= B$.

It is easy to check that the anomaly conditions of the linear model imply that the hybrid model is also anomaly-free.
Hence, we conclude that in our class of models the locus in the large radius limit corresponding to complex structure singularities has a natural interpretation
in terms of a (0,2) hybrid model with data as above,
confirming our claim that such locus gives rise to a non-singular CFT.
More importantly perhaps, the parameter space of such hybrid model
admits, by construction, a sublocus possessing the full symmetry of the GP model.
Within this sublocus the symmetries defined by $H$ are therefore geometrical, 
and the mirror quotient procedure has a well-defined action. 

In most cases, the hybrid geometry will still be singular in the sense that the base $B$ is
generically a weighted projective space and therefore possesses orbifold singularities. 
We are familiar with this type of singularities from the context of compact NLSMs and, as mentioned above, they do no introduce any particular conceptual issue. 
Thus, we will mostly ignore such singularities in the hybrid models in the next section when we study some explicit examples.
Moreover, it is often the case that it is possible to resolve these singularities and obtain a smooth geometry, though such a geometry might not admit 
a simple presentation. The details of this depend however on the data of the specific model.

It is interesting to consider the relation between the hybrid model and the more general CICY. A {\it good} hybrid model (as opposed to {\it bad} \cite{Bertolini:2013xga} 
or {\it pseudo} \cite{Aspinwall:2009qy}) is properly 
described in the hybrid limit, where the 
K\"ahler form of base $B$ is taken to be deep into its K\"ahler cone. 
In terms of our linear model, this indeed corresponds to taking $r\rightarrow\infty$. 
In this limit, where worldsheet instantons wrapping rational curves in $B$ are suppressed, the hybrid model provides an UV description of the 
sublocus of the moduli space parametrized by 
the (0,2) superpotential $\bJ\in\Gamma(\cE^\ast)$. These parameters 
have natural representatives in terms of couplings in the UV hybrid Lagrangian \cite{Bertolini:2017lcz},
and the corresponding CFT deformations respect the geometric structure of the fibration of the hybrid model.
The IR CFT admits other types of deformations. These correspond, from the UV hybrid point of view, 
to variations of the complex structure of $\bY$ and deformations of the holomorphic bundle $\cE$. 
That is, such deformations necessarily break the geometric structure of the hybrid model, 
 and turning on a subset of these generates in our case the transition between the hybrid model and the more general CICY.

\section{Examples}
\label{s:examples}

In this section we study a number of examples illustrating our mirror duality. 
In each case, we discuss the geometric realization of the theories and its relation with the mirror map. 

We proceed from a well-known example as a check of our techniques to 
more complicated models displaying a number of rather surprising conceptual features. 
For instance, we will see that there may be several geometric interpretations of the same theory, and that our
mirror map is well-defined on all of these. We shall also see that in some cases 
the target space satisfies $\dim B>\cb/3$, which, despite seemingly leading to a contradiction for a unitary theory,
is nonetheless consistent within our hybrid formalism.

\subsection{$\P^5_{111122}[4,4]$}

We start with the much-studied example \cite{Distler:1993mk} describing a rank 4 bundle over a Calabi-Yau 
complete intersection of two quartics in $V=\P^5_{111122}$. 
Following the prescription of the previous section, we construct the GLSM specified by the data
\begin{align}
\label{eq:DK44}
\xymatrix@R=1mm@C=3mm{
		&X_1	&X_2	&X_3		&X_4	&Y_1	&Y_2	&P 	&\Gamma^1	&\Gamma^2	&\Gamma^3	&\Gamma^4	&\Gammat	&\Lambda^1	&\Lambda^2	&\text{F.I.}\\
\GU(1)	&1		&1			&1			&1		&2		&2		&-5	&1			&1			&1			&1			&1			&-4			&-4	&r
}
\end{align}
In the $r\ll0$ phase
we recover the $\Z_5$ orbifold of the GP model $A_3^{\oplus4}\oplus\GP_{2,2}$.
According to our construction, the mirror dual theory is generated by a $H=\Z_5^3$ quotient. 
We list in table \ref{t:DK44GPlist} the various theories corresponding to all the quotients by subgroups of $H$, 
and observe that they organize in mirror pairs, where the $\rep{16}$ and $\rep{\overline{16}}$ representations
of $\so(10)$ are interchanged while the number of $\rep{10}$'s and gauge singlets is invariant. Note that our results
match and extend those of \cite{Blumenhagen:1996tv}.

In the $r\gg0$ phase the theory flows, for generic values of the (0,2) superpotential, to a NLSM
on $\cV\rightarrow M$ where $M=\{(x,y)\in V| H_1=H_2=0\}$ is described in terms of two quartics in $\P^5_{111122}$
and $\cV$ is defined by the restriction to $M$ of the SES
\begin{align}
\xymatrix@R=0mm@C=12mm{
0	\ar[r]	&\cV	\ar[r]	&\cO(1)^{\oplus5} \ar[r]^-{J_I}	&\cO(5)	  \ar[r]	&0~,
}
\end{align}
where $J_I$, $I=1,\dots,5$ are homogeneous polynomials of degree 4 in $x,y$. 
Comparing to the notation of the previous section, we redefined $J_5=\Jt$.
The generic theory breaks all of $H$, whose elements therefore do not correspond to  
geometrical symmetries from the point of view of the complete intersection. 
As described in section \ref{ss:hybrids}, the theory restricted to its $H$-preserving locus
is described instead in terms of the hybrid model $\cE\rightarrow\bY$, where 
\begin{align}
\label{eq:DK44hybr}
\bY&=\tot\left( \cO(-5)\oplus\cO(2)^{\oplus2} \rightarrow \P^3\right)~,\nonumber\\
\cE&=\cO(1)^{\oplus5}\oplus\cO(-4)^{\oplus2}~.
\end{align}
This model is equipped with a non-trivial superpotential which, reintroducing explicitly the $p$ dependence, satisfies
\begin{align}
\begin{pmatrix}
pF^i_{[4]}\\
py_1^2\\
y_1^2+y_2^2\\
y_1y_2
\end{pmatrix} \in \Gamma(\cE^\ast)~,
\end{align}
where $F^i_{[4]}=x_i^4\in H^0(\P^3,\cO(4))$. 

\begin{table}[t!]
\begin{center}
\begin{tabular}{|c|c|c|c|c|}
\hline
symmetries		&$\rep{\overline{16}}$	&$\rep{16}$	&$\rep{10}$		&$\rep{1}$\\
\hline
-                                &80                        		&0               	&74              		&350\\[1.5mm]
$[1,0,0,0,4]$             &36                        		&8              	&44              		&302\\[1.5mm]
$[1,2,3,4,0]$             &20                        		&4               	&26               		&230\\[1.5mm]         
$[1,1,3,0,0]$          	&18        		    	        &14              	&32               		&254\\ [1.5mm]
$[0,0,1,1,3]$             &4                        		&36              	&42               		&270\\[1.5mm]
$[1,4,0,0,0]$             &42                         		&2              	&42               		&270\\[1.5mm]
${\begin{matrix}
[1,0,0,0,4]\\ 
[0,1,0,0,4]
\end{matrix}}$		&2					&42			&42				&270\\[3.5mm]
${\begin{matrix}
[0,1,2,3,4]\\ 
[1,4,0,0,0]
\end{matrix}}$		&36					&4			&42				&270\\[3.5mm]
${\begin{matrix}
[0,1,2,3,4]\\ 
[1,0,0,0,4]
\end{matrix}}$		&14					&18			&32				&254\\[3.5mm]
${\begin{matrix}
[0,1,3,1,0]\\ 
[0,1,1,0,3]
\end{matrix}}$		&4					&20			&26				&230\\[3.5mm]
${\begin{matrix}
[0,1,0,4,0]\\ 
[0,0,1,4,0]
\end{matrix}}$		&8					&36			&44				&302\\[3.5mm]
${\begin{matrix}
[1,0,0,4,0]\\ 
[0,1,0,4,0]\\
[0,0,1,4,0]
\end{matrix}}$			&0				&80			&74				&350\\\hline
\end{tabular}
\caption{Orbifolds of the theory $A_3^{\oplus4}\oplus\GP_{2,2}/\Z_{5}$.}
\label{t:DK44GPlist}
\end{center}
\end{table}

Let us now consider the mirror model. This is given as a $H=\Z_5^3$ orbifold of the corresponding hybrid geometry.
A feature of this example is that for any mirror pair in table \ref{t:DK44GPlist}, the quotient by $H$ can always be chosen such that
it acts only on the base coordinates. 
In the case where we take the mirror of our original theory \eqref{eq:DK44hybr} we obtain the hybrid model $\cE^\circ\rightarrow\bY^\circ$, where
\begin{align}
\label{eq:DK44hybrmirr}
\bY^\circ&=\tot\left( \cO(-5)\oplus\cO(2)^{\oplus2} \rightarrow \P^3/H\right)~,\nonumber\\
\cE^\circ&=\cO(1)^{\oplus5}\oplus\cO(-4)^{\oplus2}~.
\end{align}
Although this model exhibits orbifold singularities in $B^\circ=\P^3/H$, it is nonetheless a 
well-defined hybrid model, and in principle the methods of \cite{Bertolini:2017lcz,Aspinwall:2010ve} can still be applied
to study some properties of the theory. 

In general, for each mirror pair in table \ref{t:DK44GPlist} we have the following duality between hybrid models
\begin{align}
\label{eq:DK44hybrmirrgen}
\xymatrix@R=0mm@C=12mm{
{\begin{matrix}\bY_{F}=\tot\left( \cO(-5)\oplus\cO(2)^{\oplus2} \rightarrow \P^3/F\right)\\
\cE_{F}=\cO(1)^{\oplus5}\oplus\cO(-4)^{\oplus2} 
\end{matrix}} \ar[r]&\ar[l]
{\begin{matrix}
\bY^\circ_{F}=\tot\left( \cO(-5)\oplus\cO(2)^{\oplus2} \rightarrow \P^3/F^\ast\right)\\
\cE^\circ_{F}=\cO(1)^{\oplus5}\oplus\cO(-4)^{\oplus2} \end{matrix}}
}~,
\end{align}
where $F\subset H$ and $F^\ast=F/H$.
It is natural to ask whether, for non-trivial $F$ (or $F^\ast$), the corresponding orbifold hybrid model parametrizes the singular locus of a smooth (up to 
orbifold singularities) complete intersection, as it is the case for the unorbifolded model \eqref{eq:DK44hybr}. 
It turns out that the answer is negative. Whenever the quotient is non-trivial, the
defining equations $H_{1,2}$ cannot be sufficiently generic to obtain
a transverse intersection while preserving the discrete symmetry.
However, we cannot rule out the possibility that upon resolving the orbifold singularities introduced by the quotient 
the theory admits a compact CICY phase after all. 
It would be interesting to determine under which conditions this is the case,
but we do not attempt it here.

\subsection{$\P^5_{3,5,6,10,12,12}[24,24]$ and $\P^5_{5,6,6,9,9,10}[27,18]$}

We briefly consider another set of examples for which a smooth geometry is not realized for any choice of the parameters of the (0,2) superpotential. 
Namely, starting with the Gepner model $A_{1}\oplus A_{3}^{\oplus2}\oplus A_{4}\oplus A_8$
we can generate a $\so(10)$ GP model of the sort we studied in this work by implementing either of the replacements $A_3\rightarrow\GP_{2,2}$
and $A_8\rightarrow\GP_{3,3}$, supplemented by a $\Z_{30}$ orbifold. 
We refer to the corresponding models as $\mathsf{M}_{2,2}$ and $\mathsf{M}_{3,3}$.
As far as symmetries and quotients are concerned, we can treat both models at once. We have $G=(\Z_3\times\Z_5^2\times\Z_6\times\Z_{10})/\Z_{30}$
and $H=\Z_5$. Thus, both models admit only one additional orbifold theory, which generates the mirror dual. 
We present these in table \ref{t:DK2424GPlist}.

The corresponding geometries are described by rank four bundles over the putative CICYs $\P^5_{3,5,6,10,12,12}[24,24]$ and $\P^5_{5,6,6,9,9,10}[27,18]$, 
namely complete intersections of degree $24,24$ and $27,18$ in the respective weighted projective spaces. 
As anticipated above, both models admit no choice of defining equations such that the hypersurfaces intersect transversely. 
Thus, there is actually no smooth Calabi-Yau interpretation for these models. The hybrid description for the GP superpotentials, however,
is not affected by this. The resulting models are
\begin{align}
\bY_{24,24}&=\tot\left( \cO(-30)\oplus\cO(12)^{\oplus2} \rightarrow \P^3_{3,5,6,10}\right)~,\nonumber\\
\cE_{24,24}&=\cO(3)\oplus\cO(5)\oplus\cO(6)^{\oplus2}\oplus\cO(10)\oplus\cO(-24)^{\oplus2}~,
\end{align}
and
\begin{align}
\label{eq:2718mirr}
\bY_{27,18}&=\tot\left( \cO(-30)\oplus\cO(9)^{\oplus2} \rightarrow \P^3_{5,6,6,10}\right)~,\nonumber\\
\cE_{27,18}&=\cO(3)\oplus\cO(5)\oplus\cO(6)^{\oplus2}\oplus\cO(10)\oplus\cO(-27)\oplus\cO(-18)~.
\end{align}
Let us have a quick look at the mirror geometries. Note that while for the dual of \eqref{eq:2718mirr} we have 
\begin{align}
\bY_{27,18}^\circ=\bY_{27,18}/H=\tot(X_{27,18}\rightarrow \P^3_{5,6,6,10}/H)~,
\end{align}
that is, the quotient acts only on the base and the mirror manifold is again described in terms of a hybrid geometry, 
this property does not hold for the mirror of $\mathsf{M}_{2,2}$, for which instead
\begin{align}
\bY_{24,24}^\circ=\bY_{24,24}/H=\tot\left(\cO(-30)\oplus\left(\cO(12)^{\oplus2}\rightarrow \P^3_{3,5,6,10}\right)/H\right)~.
\end{align}
In this case, the mirror theory can be interpreted solely in terms of an orbifold of a hybrid model, and not as a hybrid theory itself.

\begin{table}[t!]
\begin{center}
\begin{tabular}{|c|c|c|c|c|c|}
\hline
model								&symmetry			&$\rep{\overline{16}}$	&$\rep{16}$	&$\rep{10}$		&$\rep{1}$\\\hline
\multirow{2}{*}{$\mathsf{M_{2,2}}$}			&-                               	&34                        		&10               	&42              		&264\\[1mm]
									&$[0,1,4,0,0]$            	&10                        		&34              	&42              		&264\\[0.5mm]\hline
\multirow{2}{*}{$\mathsf{M_{3,3}}$}			&-                               	&30                        		&6               	&34              		&276\\[1mm]
									&$[0,1,4,0,0]$            	&6                        		&30              	&34              		&276\\
\hline        
\end{tabular}
\caption{Orbifolds of the theories $\mathsf{M}_{2,2}$ and $\mathsf{M}_{3,3}$.}
\label{t:DK2424GPlist}
\end{center}
\end{table}

\subsection{$\P^7_{12233445}[6,6,6,6]$}

We conclude this section with a model giving rise to a $\su(5)$ theory. 
We study the model $A_1^{\oplus2}\oplus A_7\oplus\GP_{2,4}^{\oplus2}$
supplemented by a $\Z_9$ orbifold. 
In particular, $H=\Z_3^2\times\Z_9$ and we collect various quotients by subgroups of $H$ 
in table \ref{t:DK6666GPlist}. For each of these we list in the last column the symmetries that lead to the mirror model, whose
spectrum is obtained by interchanging $\rep{10}\leftrightarrow\rep{\overline{10}}$ and $\rep{5}\leftrightarrow\rep{\overline{5}}$.
Following our discussion in the previous section, we realize the model as a phase of the GLSM determined by the following data
\begin{align}
\label{eq:SU5GLSM1}
\xymatrix@R=1mm@C=3mm{
		&X_1	&X_{2,3}		&Y_{1j}		&Y_{2j}	&Z		&P 	&\Gamma^1	&\Gamma^{2,3}	&\Gammat^{1,2}	&\Gammah	&\Lambda^1_j	&\Lambda^2_j\\
\GU(1)	&1		&3			&2			&4		&9		&-9	&1			&3				&1				&0			&-8	&-6
}
\end{align}
where $j=1,2$. At $r\gg0$ this theory describes a putative CICY
$\P^7_{12233449}[8,8,6,6]$ equipped with a rank five bundle. 
The manifold exhibits complex structure singularities at the point $x=y=0$ as the auxiliary coordinate $z$ of weight 9 is not allowed to appear in the defining 
equations of the hypersurfaces. 
The theory defined by the GP superpotential admits instead a well-defined hybrid description, with data
\begin{align}
\bY_1&=\tot\left( \cO(-9)\oplus\cO(2)^{\oplus2}\oplus\cO(4)^{\oplus2} \rightarrow \P^3_{1,3,3,9}\right)~,\nonumber\\
\cE_1&=\cO\oplus\cO(1)^{\oplus3}\oplus\cO(3)^{\oplus2}\cO(-8)^{\oplus2}\oplus\cO(-6)^{\oplus2}~.
\end{align}

\begin{table}[t!]
\begin{center}
\begin{tabular}{|c|c|c|c|c|c|c|}
\hline
symmetries	&$\rep{10}$		&$\rep{\overline{10}}$		&$\rep{5}$		&$\rep{\overline{5}}$		&$\rep{1}$	&mirror symm.\\
\hline
-			&50				&2						&98				&50					&326			&${\begin{matrix}
[0,0,0,1,8]\\
[1,0,6,0,0]\\
[0,1,0,6,0]
\end{matrix}}$\\[0mm]\hline
$[1,2,0,0,0]$	&50				&2						&98				&50					&326			&${\begin{matrix}
[0,0,1,8,0]\\
[1,1,3,0,0]
\end{matrix}}$\\\hline
$[1,0,6,0,0]$	&27				&3						&68				&44					&294			&${\begin{matrix}
[0,0,0,1,8]\\
[1,0,6,0,0]
\end{matrix}}$\\[0mm]\hline
$[1,0,0,6,0]$	&23				&5						&66				&48					&296			&${\begin{matrix}
[0,0,1,8,0]\\
[1,0,0,0,6]
\end{matrix}}$\\[0mm]\hline
$[0,0,3,6,0]$	&18				&6						&54				&42					&270			&${\begin{matrix}
[1,0,6,0,0]\\
[1,0,0,6,0]\\
[1,2,0,0,0]
\end{matrix}}$\\[0mm]\hline
$[1,1,3,0,0]$	&24				&12						&60				&48					&262			&${\begin{matrix}
[1,2,0,0,0]\\
[0,0,0,1,8]
\end{matrix}}$\\[0mm]\hline
$[1,1,0,3,0]$	&17				&11						&47				&41					&278			&${\begin{matrix}
[1,2,0,0,0]\\
[0,0,1,0,8]
\end{matrix}}$\\[0mm]\hline
${\begin{matrix}
[1,2,0,0,0]\\
[1,0,6,0,0]
\end{matrix}}$	&24				&12						&60				&48					&262			&$[0,0,0,1,8]$\\\hline
${\begin{matrix}
[1,2,0,0,0]\\
[1,0,0,6,0]
\end{matrix}}$	&17				&11						&47				&41					&278			&$[0,0,1,8,0]$\\\hline
${\begin{matrix}
[1,2,0,0,0]\\
[0,0,3,6,0]
\end{matrix}}$	&18				&6						&54				&42					&270			&${\begin{matrix}
[1,0,3,0,6]\\
[1,1,0,3,0]
\end{matrix}}$\\\hline
${\begin{matrix}
[1,0,6,0,0]\\
[0,0,0,3,6]
\end{matrix}}$	&9				&9						&44				&44					&286			&${\begin{matrix}
[0,0,3,0,6]\\
[1,0,0,3,3]
\end{matrix}}$\\\hline
${\begin{matrix}
[1,0,6,0,0]\\
[0,1,0,6,0]
\end{matrix}}$	&17				&5						&55				&43					&268			&$[1,0,7,8,0]$\\\hline
${\begin{matrix}
[1,0,0,6,0]\\
[0,1,0,0,6]
\end{matrix}}$	&12				&12						&52				&52					&250			&$[1,0,8,7,0]$\\\hline
\end{tabular}
\caption{Orbifolds of the theory $A_1^{\oplus2}\oplus A_7\oplus\GP_{2,4}^{\oplus2}/\Z_{9}$.}
\label{t:DK6666GPlist}
\end{center}
\end{table}

It is in fact possible to improve our geometric interpretation of the model by applying a target space duality \cite{Distler:1995bc,Chiang:1997kt}.
Let us consider the following inequivalent linear model
\begin{align}
\label{eq:SU5GLSM2}
\xymatrix@R=1mm@C=3mm{
		&X_1	&X_{2,3}		&Y_{1j}		&Y_{2j}	&Z		&P 	&\Gamma^{1,2,3}	&\Gammat^{1,2}	&\Gammah	&\Lambda^A\\
\GU(1)	&1		&3			&2			&4		&5		&-9	&1				&1				&4			&-6
}
\end{align}
where $A=1,\dots,4$ and we have appropriately relabeled some of the Fermi fields.
The theory describes, in its large radius phase, a rank five bundle over the CICY $\P^7_{12233445}[6,6,6,6]$.
For a sufficiently generic choice of parameters this manifold does in fact avoid complex structure singularities. A choice that
realizes this is
\begin{align}
H_A=x_1^6+x_2^2+x_3^2+y_{11}^3+y_{12}^3+ z x_1 + y_{11}y_{21}+ y_{12}y_{22}~,
\end{align}
upon taking generic coefficients. 

At $r\ll0$ instead we recover the GP model by taking the superpotential
\begin{align}
\label{eq:GP6666pot}
J_1&=x_1^8~,		&J_2&=y_{11}^2y_{12}~,	 &J_3&=y_{12}^2~,		&J_4&=y_{21}^2y_{22}~,		&J_5&=y_{22}^2~,	&J_6&=z~,\nonumber\\
H_1&=x_2^2~,		 &H_2&=x_3^2~,		&H_3&=y_{11}^3~,		&H_4&=y_{21}^3~,
\end{align}
and using the fact that at the GP point a field redefinition of the $\GP_{2,4}$ model relates the two ideals
\begin{align}
\bJ_{2,4}=\left( y_1^4,y_2^2,y_1y_2\right) \quad \longleftrightarrow \quad \bJ'_{2,4}=\left( y_1^2y_2,y_2^2,y_1^3\right)~,
\end{align}
which therefore lead to isomorphic IR conformal field theories. 
This field redefinition, trivial at the GP point, acts non-trivially in the geometric phase. In fact, the choice \eqref{eq:GP6666pot} leads 
to the non-singular hybrid model at large radius 
\begin{align}
\bY_2&=\tot\left( \cO(-9)\oplus\cO(3)^{\oplus2}\oplus\cO(4)^{\oplus2} \rightarrow \P^3_{1225}\right)~,\nonumber\\
\cE_2&=\cO\oplus\cO(1)^{\oplus5}\oplus\cO(4)\oplus\cO(-6)^{\oplus4}~,
\end{align}
while if we chose to describe any of the two $\GP_{2,4}$ component theories by the ideal $\bJ_{2,4}$ the 
corresponding geometry would have been singular.

\begin{table}
\begin{center}
\begin{tabular}{|c|cccccc|cccc|c|c|}
\hline
$q_Z$	&\multicolumn{6}{c|}{$Q_I^{\Gamma}$} 	&\multicolumn{4}{c|}{$Q^{\Lambda}_A$}	&$B$	 \\\hline
13		&1		&3		&3		&3		&3		&$-4$	&$-8$			&$-8$			&$-8$			&$-8$	&$\P^3_{1,3,3,13}$	\\\hline
11		&1		&1		&3		&3		&3		&$-2$	&$-8$			&$-8$			&$-8$			&$-6$	&$\P^4_{2,3,3,4,11}$\\\hline
9		&1		&1		&1		&3		&3		&0		&$-8$			&$-8$			&$-6$			&$-6$	&$\P^3_{1,3,3,9}$	\\\hline
7		&1		&1		&1		&1		&3		&2		&$-8$			&$-6$			&$-6$			&$-6$	&$\P^3_{1,2,4,7}$\\\hline
5		&1		&1		&1		&1		&1		&4		&$-6$			&$-6$			&$-6$			&$-6$	&$\P^3_{1,2,2,5}$\\\hline

\end{tabular}
\caption{Gauge charges of left-moving fermions determining rank-5 bundles over $\P^7_{1223344q_Z}[-Q^\Lambda_A]$.}
\label{t:SU5hybrlsist}
\end{center}
\end{table}

Other non-anomalous possibilities are realized for different values of $q_Z$,
the gauge charge of the field $Z$. We report these in table \ref{t:SU5hybrlsist},
where we also list the base $B$ of the hybrid target space geometry for a choice of superpotential interactions
that respects the discrete symmetry group $H$ of the GP model.
It is worth noting that in one case the space of classical vacua $B$ is four-dimensional.
Let us look in some detail at the example where this phenomenon occurs. 
This is the model with $q_Z=11$ and superpotential
\begin{align}
H_1&=x_1^8~,		 	&H_2&=y_{11}^4~,		&H_3&=y_{12}^2~,			&H_4&=y_{11}y_{12}~.
\end{align}
leading to
\begin{align}
\bY_3&=\tot\left(\cO(-9)\oplus\cO(1)\oplus\cO(2)\oplus\cO(4)\rightarrow \P^4_{2,3,3,4,11}\right)~,\nonumber\\
\cE_3&=\cO(1)^{\oplus2}\oplus\cO(3)^{\oplus3}\oplus\cO(-2)\oplus\cO(-8)^{\oplus3}\oplus\cO(-6)~.
\end{align}
Naively, one would imagine that the right-moving central charge receives contributions from the NLSM describing the $D=4$-dimensional 
base ($\cb=12$) and the fiber LG theory ($\cb_{\text{LG}}\geq0$), in contradiction with our claim that this theory flows to a CFT with $\cb=9$. 
The expression for the central charge of a (0,2) hybrid model \cite{Bertolini:2017lcz} yields however the expected result
\begin{align}
\cb = 3 \left(D +r+n-R\right)=3 \left(4+5+4-10 \right)=9~,
\end{align}
where $n=\dim\bY-\dim B=4$, $R=\rank\cE_3=10$ and $r=5$ is the level of the left-moving $\mathfrak{u}(1)_L\subset \su(5)$ current. 
Thus, it may appear that the contribution to the central charge of the fiber LG theory is $\cb_{\text{LG}}=-3$, which is forbidden for a unitary theory. 
The resolution to the puzzle lies in the invalid assumption that the full hybrid theory, although it can be thought of
as a LG fibration over a compact base, requires a fiberwise well-defined LG theory. In this example, the fiberwise theory is not
well-defined as a standalone Landau-Ginzburg theory, 
due to the fact that LG models develop singularities whenever the the R-charge of some bosonic field vanishes. 
Such decompactifications are not present in the full hybrid model due to the global constraints
introduced by the nontrivial structure of the fibration of $\bY_3$.

\section{Discussion}
\label{s:outlook}

We have described a mirror duality for a large class of (0,2) models without a (2,2) locus.
The duality is realized as a quotient at an exactly solvable point in the moduli space of
CICYs with bundles whose rank is
strictly greater than that of the tangent bundle. 
In this final section, we speculate upon the implications of such a mirror duality on the structure of the 
moduli space of (0,2) SCFTs, and discuss several lines of investigations worth detailed study.
In order to keep the discussion a bit more concrete, we will mainly restrict ourselves here to the case $\cb=9$.

The explicit map of our duality involves, in particular, the interchange of the matter content of the theories, that is, operators charged under the spacetime gauge group.  
On the one hand, in the (0,2) setting there is no a priori relation between these and the uncharged operators,
thus it is unclear how to infer the effect of the mirror map on the moduli of the theory.
On the other hand, this interchange nevertheless implies an isomorphism between the A/2 subring of the chiral ring of the (0,2) theory
and the B/2 subring of the mirror model. 
Recent advances in the development of techniques to compute correlators in the corresponding twisted theories \cite{Bertolini:2018now,Closset:2015ohf}
are likely to shed light on the moduli dependence on at least some subsectors of the theory. 
For instance, some indications that a splitting between K\"ahler and complex structure moduli
occurs already appeared in the class of (0,2) theories studied in \cite{Melnikov:2010sa,Bertolini:2018qlc}. 
It would be worthwhile to determine also in our models whether at least 
some coarse quantities, as for instance the discriminant loci of the twisted theories, are indipendent
of some subset of the parameters of the theory.
One may then more generally test whether the entire set of B/2 model correlators 
does not suffer from worldsheet instanton corrections. If true, even for a subclass of theories, the mirror map would then allow
the computation of a set of quantum corrected objects in terms of classical quantities in the mirror theory.

As pointed out in the main body of the text, the effectiveness in exploring the moduli space of our models is impaired by
our limited understanding of resolutions of singularities in the context of (0,2) theories. 
If mirror symmetry is indeed a property of the (0,2) moduli space that goes beyond LG orbifolds, 
there should be a combinatorial structure that generalizes our construction. 
Experience with (2,2) theories \cite{Aspinwall:2015zia} (and deformations thereof)
suggests the natural place to begin such a search is the gauged linear sigma model. 

In this context, our results showed that the concept of phases of linear models must be somehow generalized. In fact, we have described how a
perturbative transition between good hybrid models and NLSMs on a compact CICY 
may occur within the same GLSM phase. Thus, it appears that in the (0,2) setting, in addition to the familiar K\"ahler phase structure determined by the
F.I.~parameters, there are also complex structure phase transitions. 
It would be interesting to determine which form such complex structure phase transitions assume in the mirror model.

Besides further investigating the theories we constructed, it would be equally important to increase the set
of models amenable to such computations.  
Although the number of theories we can generate with our approach is quite large, 
in the landscape of (0,2) compactifications the models considered in this work and to which our methods apply are all but generic. 
In fact, the existence of a Landau-Ginzburg description as well as of a (0,2) NLSM phase  
are both highly non-trivial conditions for a (0,2) heterotic background. 
Even in the realm of (0,2) Landau-Ginzburg models, our theories comprise only a subset of it. How large this subset is remains 
hitherto unknown, as we lack a complete classifications of (0,2) LG theories even at fixed central charge, 
in contrast to the (2,2) case \cite{Kreuzer:1992bi,Kreuzer:1992np,Davenport:2016ggc}.
However, recent progress in this direction \cite{Gholson:2018zrl} seems to hint that an analogous version of the mirror duality holds beyond the class of models
under study in our present work.

\appendix

\section{Characters conventions}
\label{app:charconvs}

In this appendix we collect our conventions for the characters of the various systems appearing in the main body of the text together with their 
modular transformations properties. 

\subsection*{Parafermionic characters}

The characters for the parafermionic theories \cite{Fateev:1985mm}	at any level $k$, which we denote $\text{PF}_k$,
have been derived in \cite{Gepner:1986hr} in terms of Hecke indefinite modular forms, exploiting the relation between the parafermionic theory and the 
$\SU(2)$ WZW model. These exhibits the following modular transformations
\begin{align}
\label{eq:PFchtransf}
\chi^{\text{PF}_k}_{\alpha,\nu} (\tau+1)&=\exp\left[ 2\pi i\left(\frac{\alpha(\alpha+2)}{4(k+2)}-\frac{\nu^2}{4k}-\frac{c_{\text{PF}_k}}{24}\right)\right]\chi^{\text{PF}_k}_{\alpha,\nu} (\tau)~,\nonumber\\
\chi^{\text{PF}_k}_{\alpha,\nu} \left(-\frac1\tau\right)&=\frac1{\sqrt{k(k+2)}}  \sum_{\alpha'=0}^k \sum_{\nu'\in\Z_{2k}} e^{i\pi \frac{mm'}{k}}\sin \frac{\pi(\alpha+1)(\alpha'+1)}{k+2}\chi^{\text{PF}_k}_{\alpha',\nu'}(\tau)~,
\end{align}
where
\begin{align}
c_{\text{PF}_k}&=2\frac{k-1}{k+2}~,
\end{align}
is the central charge of the $\text{PF}_k$ model. We implement the selection rule of the parafermionic theory by
defining $\chi^{\text{PF}_k}_{\alpha,\nu} \equiv0$ whenever $\alpha+\nu=1\mod2$.

\subsection*{$\GU(1)^2$ characters}

Here we consider the characters for the $\GU(1)^2$ current algebra with anomaly matrix given by
\begin{align}
Q=\begin{pmatrix}
m^2	&1\\
1	&n^2
\end{pmatrix}~.
\end{align}
We define the characters
\begin{align}
\chi^{\GU(1)_Q^{2}}_{\br-\frac{\bbb}2,t}(\tau,\bxi)&=\eta^{-2} \sum_{\substack{\blambda\in\Z^{2}\\m\lambda_1+n\lambda_2=t\mod2}} 
e^{\pi i \tau \left(\blambda+(\br-\frac{\bbb}2) Q^{-1} \right)Q\left(\blambda+(\br-\frac{\bbb}2) Q^{-1} \right)^\top+2\pi i(\blambda Q+(\br-\frac{\bbb}2) ) \bxi^\top}~,
\end{align}
where $\br\in\Z^2$ and
\begin{align}
\bbb=b\begin{pmatrix}
m	&n
\end{pmatrix}~,	\qquad\qquad b\in\Z_2~.
\end{align}
We note that these are not characters of irreducible representations
of the current algebra but of collections of such representations,
indexed by $\lambda_{1,2}$.
These transform under $T$ and $S$ as follows
\begin{align}
\label{eq:U12chtransf}
\chi^{\GU(1)_Q^{2}}_{\br-\frac{\bbb}2,t}(\tau+1,\bxi)&=e^{-2\pi i\frac2{24}}e^{2\pi i\left(\half(\br-\frac{\bbb}2)Q^{-1}(\br-\frac{\bbb}2)^\top+\frac{(1-b)t}2\right)}
\chi^{\GU(1)_Q^{2}}_{\br-\frac{\bbb}2,t}(\tau,\bxi)~,\nonumber\\
\chi^{\GU(1)_Q^{2}}_{\br-\frac{\bbb}2,t}(-\frac1{\tau},\frac{\bxi}{\tau})&=\frac1{2\sqrt{\det Q}}
e^{\frac{\pi i}{\tau}\bxi Q \bxi^\top} \sum_{\br'\in\Z^2_Q}\sum_{b'\in\Z_2}\sum_{t'\in\Z_2}e^{-2\pi i \left( (\br'-\frac{\bbb'}2) Q^{-1}(\br-\frac{\bbb}2)^\top+\frac{bt'+tb'}2 \right)} \chi^{\GU(1)^{2}_Q}_{\br'-\frac{\bbb'}2,t'}(\tau,\bxi)~.
\end{align}
The notation $\Z^2_Q\equiv \Z^2/Q\Z^2$ defines the lattice for the $\GU(1)^2$ charges.

\subsection*{$N=2$ minimal model characters}

For completeness, we also present the transformations of the characters of the $N=2$ minimal models \cite{Qiu:1987ux}. 
In particular, since in our conventions the right-moving side is supersymmetric, we list the modular transformations for the anti-holomorphic characters, which read
\begin{align}
\chib^{\alphab}_{\qb,\sb}(\tau+1)&=\exp\left[ -2\pi i\left(\frac{\alphab(\alphab+2)}{4(k+2)}-\frac{\qb^2}{4(k+2)}+\frac{\sb^2}{8}-\frac{\cb_k}{24}\right)\right] 
\chib^{\alphab}_{\qb,\sb}(\tau)~,\nonumber\\
\chib_{\qb,\sb}^{\alphab} \left(-\frac1\tau\right)&=\frac1{2(k+2)}  \sum_{\alphab'+\qb'+\sb'=0\mod2} e^{2i\pi \left(-\frac{\qb\qb'}{2(k+2)} +\frac{\sb\sb'}4\right)}\sin \frac{\pi(\alphab+1)(\alphab'+1)}{k+2}\chib_{\qb',\sb'}^{\alphab'}(\tau)~,
\end{align}
where $\cb_k=3k/(k+2)$ is the central charge.

\subsection*{$\SO(2k)$ characters}

We consider here the level one $\widehat{\so(2k)}$ current algebra
realized, for example, by $2k$ free chiral fermions.  
The integrable representations at level one for these groups are the singlet, vector and the spinor/anti-spinor. We group them in a vector
\begin{align}\label{eq:Blambdas}
\bB_{\lambda}=\begin{pmatrix}
B_{\text{o}}	&B_{\text{v}}		&B_{\text{s}}		&B_{\overline{\text{s}}}
\end{pmatrix}~.
\end{align}
These transform under the modular group according to the matrices
\begin{align}
T_{2k}&=e^{-2\pi i\frac{k}{24}}\begin{pmatrix}
1	&0	&0					&0\\
0	&-1	&0					&0\\
0	&0	&e^{\frac{\pi i k}4}	&0\\
0	&0	&0					&e^{\frac{\pi ik}4}
\end{pmatrix}~,
&S_{2k}&=\half \begin{pmatrix}
1	&1	&1			&1\\
1	&1	&-1			&-1\\
1	&-1	&e^{-\pi i\frac{k}2 }		&-e^{-\pi i\frac{k}2 }\\
1	&-1	&-e^{-\pi i\frac{k}2 }		&e^{-\pi i\frac{k}2 }
\end{pmatrix}~.
\end{align}
We can write the above transformations as follows
\begin{align}
\label{eq:SO2ktransf}
T_{2k}&=e^{-2\pi i\frac{k}{24}} e^{2\pi i\left(\frac{r^2}8+(k-1)\frac{d^2}8\right)}~,
&S_{2k}&=\half  e^{-2\pi i\left(\frac{rr'}4+(k-1)\frac{dd'}4\right)}~,
\end{align}
where $r=0$ for the singlet, $r=2$ for the vector, $r=\pm1$ for the spinor representations and we define $d\equiv r\mod2$.
We can also choose the following basis
\begin{align}
\bBt_{\lambda}=\half \begin{pmatrix}
B_{\text{o}}+B_{\text{v}}	&B_{\text{o}}-B_{\text{v}}		&B_{\text{s}}+B_{\overline{\text{s}}}		&B_{\text{s}}-B_{\overline{\text{s}}}
\end{pmatrix}~,
\end{align}
in terms of which the modular transformations take the form
\begin{align}
\label{eq:SO2ktilde}
\Tt_{2k}&=e^{-2\pi i\frac{k}{24}}\begin{pmatrix}
0	&1	&0					&0\\
1	&0	&0					&0\\
0	&0	&e^{\frac{\pi i k}4}	&0\\
0	&0	&0					&e^{\frac{\pi ik}4}
\end{pmatrix}~,
&\St_{2k}&= \begin{pmatrix}
1	&0	&0			&0\\
0	&0	&1			&0\\
0	&1	&0			&0\\
0	&0	&0			&i^{-k}
\end{pmatrix}~.
\end{align}

\subsection*{$\GE_8$ characters}

In our story only the singlet representation of $\GE_8$ plays a role. If we denote such representation $\chi^{\GE_8}_\text{0}$, we have
\begin{align}
T\chi^{\GE_8}_\text{0}&= e^{-2\pi i \frac{8}{24}} \chi^{\GE_8}_\text{0}~,		&S\chi^{\GE_8}_\text{0}&= \chi^{\GE_8}_\text{0}~.
\end{align}

\section{Modular invariance of $Z_{\text{proj}}$}
\label{app:modinv}

In this appendix we compute the modular transformations of the partition function 
\begin{align}
\label{eq:hetpfmodinvap}
Z_{\text{proj}}&=\frac14 \sum_{\balpha,\balphab} \sum_{\substack{\bmu,\bmub,\bnu,\by\\ 2\bgamma_a\cdot\bmu\in\Z}} N_{\balpha\balphab} e^{\pi i b_0} \chi^{\balpha}_{\bmu;\bnu}\chib^{\balphab}_{\bmu+\bmu_0+2y^a\bgamma_a}~,
\end{align} 
and thus explicitly show that it defines a modular invariant theory.\footnote{In this appendix will we often use the Einstein's summation convention.} 
The variables $y^a\in\Z$, $a=N_1+1,\dots,R$ implement the sum over the twisted sectors which are crucial, as we shall see, to recover modular invariance. 
Here, we will mostly ignore the $N=2$ minimal model components since such theories are independently modular invariant.

To show invariance under $T$ is straightforward since the conditions $2\bgamma_a\cdot\bmu\in\Z$ implies 
$b_0=b_j$ for all $a$, and 
\begin{align}
T\chi^{\balpha}_{\bmu;\bnu}\big|_{b_0=b_a}&=
 e^{\pi  i(\bmu+\bmu_0)\cdot(\bmu+\bmu_0)} T_{\balpha} e^{-2\pi i \frac{4+2d}{24}}e^{\pi i(3-d) \frac{b_0^2}4} \chi^{\balpha}_{\bmu;\bnu} ~,\nonumber\\
T\chib^{\balphab}_{\bmu+\bmu_0+2y^a\bgamma_a}\big|_{b_0=b_a}&=e^{2\pi i \frac{4+2d}{24}} \Tb_{\balphab} e^{-\pi i(\bmu+\bmu_0)\cdot (\bmu+\bmu_0)}e^{-\pi i(3-d) \frac{b_0^2}4}\chib^{\balphab}_{\bmu+\bmu_0+2y^a\bgamma_a}~.
\end{align} 
To verify invariance under $S$ is slightly more involved. We find it convenient to introduce the notation 
$s_a=2r_a-b_a$, where $s_a\in\Z_4$ implies that $r_a,b_a\in\Z_2$. The relevant transformations then read
\begin{align}
S\chi^{\balpha}_{\bmu;\bnu}\big|_{b_0=b_j}&= \sum_{\balpha',\bmu',\bnu'} S_{\balpha,\balpha'} e^{-2\pi i \bmu\cdot\bmu'}
e^{-2\pi i b_0\left((7-d)\frac{ b_0'}4+\sum_{a=N_1+1}^R\frac{b_{a}'-b_0'}4 \right)} \chi^{\balpha'}_{\bmu';\bnu'}~,\nonumber\\
S\chib^{\balphab}_{\bmu+\bmu_0+2y^a\bgamma_a}\big|_{b_0=b_a}&= \sum_{\balphab',\bmub'} S_{\balphab,\balphab'} e^{2\pi i (\bmu+\bmu_0+2y^a\bgamma_a)\cdot\bmub'}
e^{2\pi i(3-d) \frac{b_0 \bb_0'}4} \chi^{\balphab'}_{\bmub'}~.
\end{align} 
Plugging these in \eqref{eq:hetpfmodinvap} we obtain
\begin{align}
SZ_{\text{proj}}&=\frac14 \sum_{\substack{\balpha,\balphab,\bmu,\bnu,y^a\\ 2\bgamma_a\cdot\bmu\in\Z}} \sum_{\balpha',\bmu',\bnu',\balphab',\bmub'}  N_{\balpha\balphab} S_{\balpha,\balpha'}S_{\balphab,\balphab'} e^{-2\pi i \bmu\cdot(\bmu'-\bmub')} e^{2\pi i \bmu_0\cdot\bmub'}\nonumber\\
&\qquad\qquad\qquad \times e^{2\pi i y^a(2\bgamma_a\cdot\bmub')}e^{-2\pi i b_0\left((3-d)\frac{ b_0'-\bb_0'}4+\sum_{a}\frac{b_{a}'-b_0'}4-\half \right)} \chi^{\balpha'}_{\bmu';\bnu'}\chib^{\balphab'}_{\bmub'}~.
\end{align} 
Summing over the various $y^a\in\Z_2$ we have the condition 
\begin{align}
\sum_{y^a\in\Z_{2}}e^{2\pi i y^a(2\bgamma_a\cdot\bmub')}=
\begin{cases}  2 & \text{if }  2\bgamma_a\cdot\bmub'\in\Z  \\
0 &\text{otherwise} 
\end{cases}~,
\end{align}
which implies that $\bb'_0=\bb'_a$ for $a=N_1+1,\dots,R$. Thus we have
\begin{align}
SZ_{\text{proj}}&=\frac{2^{N_2}}{4} \sum_{\substack{\balpha,\balphab,\bmu,\bnu\\ 2\bgamma_a\cdot\bmu\in\Z}}  \sum_{\substack{\balpha',\bmu',\bnu',\balphab',\bmub'\\2\bgamma_a\cdot\bmu'\in\Z}}  N_{\balpha'\balphab'}S_{\balpha,\balpha'}S_{\balphab,\balphab'}
e^{-2\pi i b_0\left((3-d)\frac{ b_0'-\bb_0'}4+\sum_{a}\frac{b_{a}'-b_0'}4-\half-\frac{r_0'-\rb_0'}2+\sum_a \frac{r_a'-\rb_a'}2 \right)} \nonumber\\
&\qquad\qquad \times e^{2\pi i \bmu_0\cdot\bmub'}
e^{-2\pi i \left( -\frac{r_0(b_0'-\bb_0')}2+\frac{b_0(b_0'-\bb_0')}4+ \sum_{a}\left(-\frac{r_a(b_a'-\bb_a')}2+\frac{b_a(b_a'-\bb_a')}4\right)\right)} \chi^{\balpha'}_{\bmu';\bnu'}\chib^{\balphab'}_{\bmub'}~.
\end{align} 
Next, summing over $r_0,r_a\in\Z_2$ we similarly obtain the conditions 
$b_0'=b_a'=\bb_0'$. This simplifies the previous expression to
\begin{align}
SZ_{\text{proj}}&=\frac{2^{2N_2+1}}{4} \sum_{\balpha,\balphab,b_0,\vec\bq} \sum_{\substack{\balpha',\bmu',\bnu',\balphab',\bmub'\\2\bgamma_a\cdot\bmu'\in\Z}}  N_{\balpha'\balphab'}
S_{\balpha,\balpha'}S_{\balphab,\balphab'}e^{\pi i b_0'}e^{-2\pi i b_0\left(-\half-\frac{r_0'-\rb_0'}2+\sum_a \frac{r_a'-\rb_a'}2 \right)} \chi^{\balpha'}_{\bmu';\bnu'}\chib^{\balphab'}_{\bmub'}~,
\end{align} 
where we defined $\vec{\bq}=(q_1,\dots,q_R)$. The sum over $b_0$ yields the constraint 
\begin{align}
\rb_0'+\sum_a \rb_a'=r_0+1+\sum_a r_a'\in\Z 	\qquad\Longrightarrow \qquad 	\bmub'-\bmu' = \bmub_0+2y^a{}'\bgamma_a~.
\end{align}
Finally, the modular transformations of the $\widehat{\su(2)}$ invariants for the indices $\balpha,\balphab$ and of the theta functions for the indices $\vec\bq$ do not involve any 
additional complications, since these factorize from the $s$-index transformations, and we obtain
\begin{align}
SZ_{\text{proj}}&=\frac14 \sum_{\substack{\balpha',\bmu',\bnu',\balphab'\\2\bgamma_a\cdot\bmu'\in\Z}}  N_{\balpha'\balphab'}
e^{\pi i b_0'} \chi^{\balpha'}_{\bmu';\bnu'}\chib^{\balphab'}_{\bmu'+\bmu_0+2y^a{}'\bgamma_a}~,
\end{align} 
proving our claim.

\bibliographystyle{./utphys}
\bibliography{./bigref}

\end{document}